%% file: main.tex
\documentclass[10pt]{article}
\usepackage[letterpaper, textwidth=6in, textheight=9in, top=1in, centering]{geometry}
\usepackage[T1]{fontenc}
\usepackage{amsmath}
\usepackage{newtxtext,newtxmath}  
\usepackage{microtype}
\usepackage{graphicx}
\usepackage{booktabs}
\usepackage{array}
\usepackage{xcolor}
\definecolor{genc}{HTML}{3B6EA8}  
\usepackage{pifont}
\usepackage{tikz}
\usepackage[numbers,sort&compress]{natbib}
\usepackage[colorlinks=true, citecolor=genc, linkcolor=genc, urlcolor=genc]{hyperref}  
\usepackage{cleveref}
\usepackage{fvextra}
\usepackage{caption}
\usepackage{placeins}  

\usepackage[skip=6pt plus 1pt minus 0.5pt, indent=0pt]{parskip}
\usepackage{titlesec}
\usepackage{needspace}
\usepackage{etoolbox}
\pretocmd{\section}{\needspace{6\baselineskip}}{}{}
\pretocmd{\subsection}{\needspace{5\baselineskip}}{}{}
\pretocmd{\paragraph}{\needspace{4\baselineskip}}{}{}
\titlespacing*{\section}{0pt}{2.4ex plus .4ex minus .2ex}{1.2ex plus .2ex}
\titlespacing*{\subsection}{0pt}{2.0ex plus .3ex minus .2ex}{0.9ex plus .2ex}
\titlespacing*{\paragraph}{0pt}{1.4ex plus .3ex minus .2ex}{0.8em}
\definecolor{ink}{HTML}{24292F}
\definecolor{mute}{HTML}{6E7781}
\definecolor{okc}{HTML}{1A7F37}
\definecolor{noc}{HTML}{CF222E}
\newcommand{\yes}{\textcolor{okc}{\ding{51}}}
\newcommand{\no}{\textcolor{noc}{\ding{55}}}
\newcommand{\partly}{\textcolor{mute}{$\circ$}}
\newcolumntype{L}[1]{>{\raggedright\arraybackslash}p{#1}}
\newcolumntype{C}[1]{>{\centering\arraybackslash}p{#1}}
\usepackage{tabularx}
\newcolumntype{Y}{>{\raggedright\arraybackslash}X}
\newcolumntype{Z}{>{\centering\arraybackslash}X}
\newcommand{\mk}[2]{\tikz[baseline=(m.base)]\node[circle, fill=#1, text=white,
  inner sep=0.6pt, minimum size=3.8mm, font=\scriptsize\sffamily\bfseries] (m) {#2};}
\newcommand{\mA}{\mk{genc}{A}}
\newcommand{\mB}{\mk{genc}{B}}
\newcommand{\mC}{\mk{noc}{C}}
\newcommand{\pbar}[2]{\tikz[baseline=-0.6ex]{
  \draw[draw=ink, fill=#1, line width=0.4pt] (0,-0.13) rectangle (0.55,0.13);
  \draw[draw=ink, fill=#2, line width=0.4pt] (0.55,-0.13) rectangle (1.1,0.13);}}

\input{figures/build/data.tex}

\input{figures/build/t3_rows.tex}
\input{figures/build/t4_rows.tex}
\input{figures/build/t6_rows.tex}
\input{figures/build/t7_rows.tex}
\input{figures/build/loss_macros.tex}
\input{figures/build/wc002.tex}

\input{figures/build/wc001.tex}
\input{figures/build/bd_macros.tex}

\newcommand{\WB}{WorldBench}
\newcommand{\term}[1]{\emph{#1}}  

\makeatletter
\renewcommand{\maketitle}{%
  \begin{center}
    {\LARGE\bfseries \@title\par}
    \vskip 1.1em
    {\@author\par}
  \end{center}
  \vskip 1.2em}
\makeatother
\renewenvironment{abstract}{%
  \centerline{\large\bfseries Abstract}\vspace{0.5ex}
  \begin{list}{}{\leftmargin=0.4in \rightmargin=0.4in \listparindent=0pt \itemindent=0pt
                  \parsep=0pt}\item\relax}{\end{list}\vspace{1ex}}

\title{{\color{genc}\WB}: Evaluating LLMs on\\ Three.js Voxel World Generation}
\author{\textbf{Krish Bakshi}\\[2pt]
  \texttt{krishrajeshbakshi@gmail.com}}
\date{}

\begin{document}
\maketitle
\input{sections/00_abstract.tex}

\input{sections/01_intro.tex}

\begin{figure}[t]
  \centering
  \includegraphics[width=\textwidth]{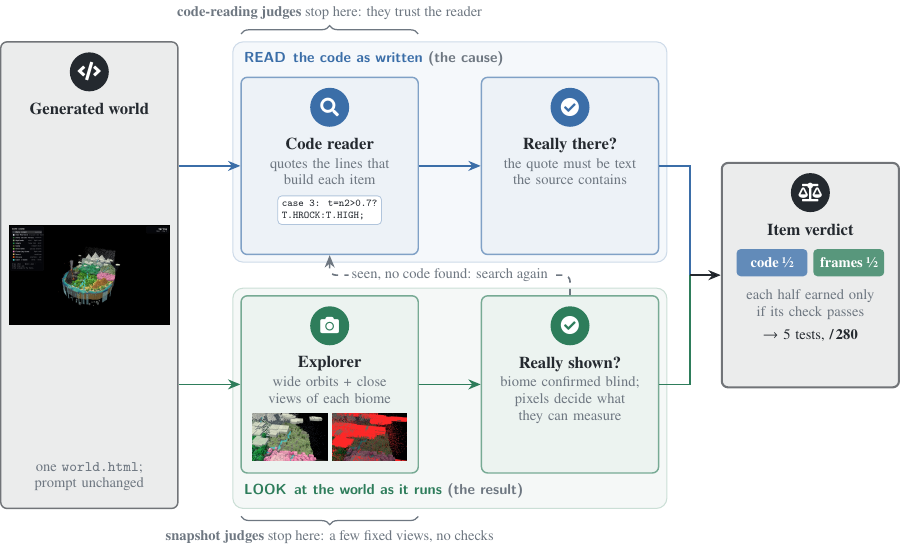}
  \caption{The judge. One generated world is examined through two channels: the
  code as written and the world as it runs. Each channel's evidence is checked
  against something the model under test cannot fake (the source text, measured
  pixels) before an item's half is earned. Grey brackets mark where code-reading and
  snapshot judges stop. Thumbnails and the code line are from \texttt{claude-fable-5-1}.}
  \label{fig:arch}
\end{figure}

\input{sections/02_related.tex}
\input{sections/03_benchmark.tex}

\begin{table}[t]
  \centering\small
  \input{figures/tab_ladder_body.tex}
  \caption{The ladder: five tests with fixed maxima, 280 points in total.}
  \label{tab:ladder}
\end{table}

\input{sections/04_method.tex}

\input{sections/05_results.tex}

\input{sections/06_limitations.tex}

\FloatBarrier
\setlength{\bibsep}{3pt plus 1pt}  
\bibliographystyle{plainnat}
\bibliography{literature/refs}

\clearpage
\appendix
\begin{center}
  {\LARGE\bfseries Appendix}
\end{center}
\addcontentsline{toc}{section}{Appendix}
\vspace{4pt}
\section{The prompt}
\label{app:prompt}
Every model receives this prompt as its only instruction (typographic characters
shown in ASCII).
{\footnotesize
\input{figures/prompt_verbatim.tex}}

\section{System and capture details}
\label{app:system}

\begin{figure}[!htb]
  \centering
  \includegraphics[width=\textwidth]{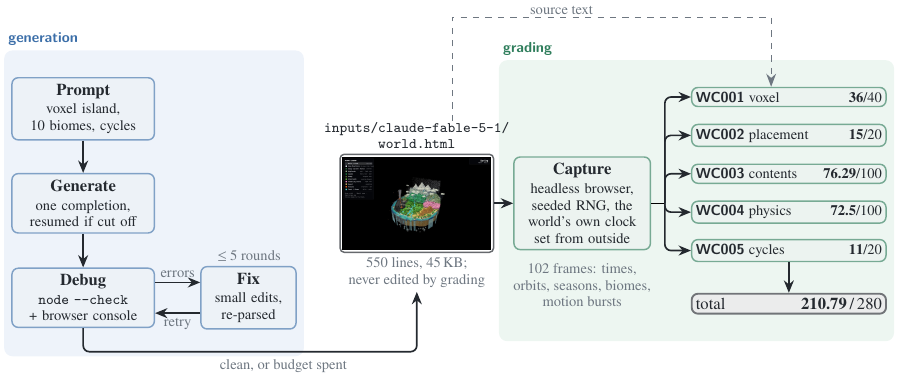}
  \caption{The full system on the reference world. Generation (left) and grading
  (right) share nothing but the world file. Scores are the reference world's real
  per-test results.}
  \label{fig:pipeline}
\end{figure}

\paragraph{Models under test.} Each world was generated by the model named below,
queried through its developer's official API with reasoning enabled where the
model supports it, at temperature 1.0 and no output-token cap. A truncated
response was continued, up to three turns in total, and the repair loop
(Section~\ref{sec:benchmark}) ran for at most five rounds. One world per model is
graded. Figure~\ref{fig:pipeline} shows the whole system, generation to score, on the
reference world.

\begin{table}[!htb]
  \centering\small
  \begin{tabular*}{\textwidth}{@{\extracolsep{\fill}}l l l l@{}}
  \toprule
  \textbf{model} & \textbf{developer} & \textbf{label in this paper} & \textbf{generated} \\
  \midrule
  Claude Fable 5.1 & Anthropic       & \texttt{claude-fable-5-1} & 2026-09-02 \\
  GPT-6 Astra      & OpenAI          & \texttt{gpt-6-astra}      & 2026-09-08 \\
  Kimi K3          & Moonshot AI     & \texttt{kimi-k-3}         & 2026-07-20 \\
  Grok 4.7         & xAI             & \texttt{grok-4-7}         & 2026-09-22 \\
  Gemini 3.1 Pro   & Google DeepMind & \texttt{gemini-3-1-pro}   & 2026-07-20 \\
  \bottomrule
  \end{tabular*}
  \caption{Models under test. All were accessed through their developers' official APIs.}
  \label{tab:models}
\end{table}

\begin{figure}[!htb]
  \centering
  \includegraphics[width=\textwidth]{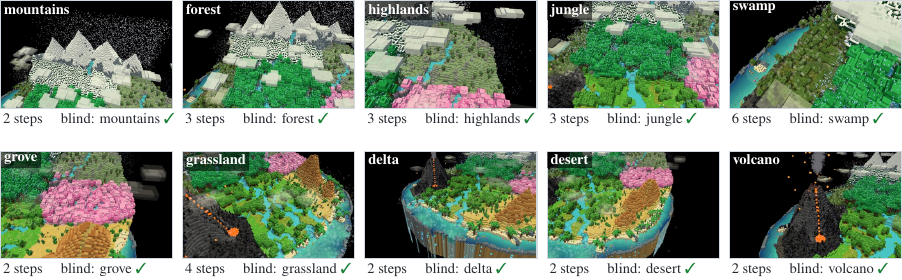}
  \caption{Every close view of the reference world: the frame each navigator episode
  saved (interface hidden), the episode's step count, and the blind identification
  answer. All ten were confirmed.}
  \label{fig:sheet}
\end{figure}

\input{sections/E_scoring.tex}
\input{sections/B_breakdown.tex}
\input{sections/C_failures.tex}
\input{sections/A_judges.tex}

\end{document}

%% file: figures/build/data.tex
\newcommand{\wbLabelA}{Exposed rock/stone}
\newcommand{\wbQuoteA}{case 3: t=n2>0.7?T.HROCK:T.HIGH;\allowbreak{} break;\allowbreak{}}
\newcommand{\wbSeenA}{``Grey stone patches on surface''}
\newcommand{\wbCodeA}{\yes}

\newcommand{\wbCodeFillA}{okc}
\newcommand{\wbVisFillA}{okc}
\newcommand{\wbEarnedA}{1.11}
\newcommand{\wbPointsA}{1.11}

\newcommand{\wbLabelB}{Mist particles}
\newcommand{\wbQuoteB}{systems.push(\allowbreak{}new PSys(\allowbreak{}\{\allowbreak{}count:600,\allowbreak{}color:0xd0d8dc,\allowbreak{}size:3,\allowbreak{}opacity:0\,\allowbreak{}\dots}
\newcommand{\wbSeenB}{\textit{not seen in any frame}}
\newcommand{\wbCodeB}{\yes}

\newcommand{\wbCodeFillB}{okc}
\newcommand{\wbVisFillB}{white}
\newcommand{\wbEarnedB}{0.56}
\newcommand{\wbPointsB}{1.11}

\newcommand{\wbLabelC}{Meltwater stream}
\newcommand{\wbQuoteC}{\textrm{\textit{no quote (probe: not in scope)}}}
\newcommand{\wbSeenC}{``Blue meltwater stream flowing down the mountain side.''}
\newcommand{\wbCodeC}{\no}

\newcommand{\wbCodeFillC}{white}
\newcommand{\wbVisFillC}{okc}
\newcommand{\wbEarnedC}{0.83}
\newcommand{\wbPointsC}{1.67}
\newcommand{\wbLineA}{163}
\newcommand{\wbLineC}{198}
\newcommand{\wbLineB}{451}

\newcommand{\wbChromaB}{0.019}
\newcommand{\wbChromaC}{0.035}
\newcommand{\wbChromaD}{0.024}
\newcommand{\wbSeasonSeen}{visuals remain identical}

\newcommand{\wbNavSteps}{6}

\newcommand{\wbRejConf}{0.7}

\newcommand{\wbRejIdent}{forest}

\newcommand{\wbBurstModelA}{claude-fable-5-1}
\newcommand{\wbBurstFracA}{38.4}
\newcommand{\wbBurstGap}{2}

\newcommand{\wbBurstModelB}{gpt-6-astra}
\newcommand{\wbBurstFracB}{0.0}
\newcommand{\wbBurstSeenB}{Vertical falling rain particles}

\newcommand{\wbBurstItem}{rain\_shaft\_fall}

%% file: figures/build/t3_rows.tex
\newcommand{\tthreerows}{%
\texttt{claude-fable-5-1} & 47 & 19 & 1 & 6 & 73 & 27\% \\
\texttt{gpt-6-astra} & 43 & 21 & 2 & 7 & 73 & 32\% \\
\texttt{kimi-k-3} & 39 & 21 & 2 & 11 & 73 & 32\% \\
\texttt{grok-4-7} & 31 & 26 & 4 & 12 & 73 & 41\% \\
\texttt{gemini-3-1-pro} & 24 & 18 & 1 & 30 & 73 & 26\% \\
\midrule
\textbf{all} & \textbf{184} & \textbf{105} & \textbf{10} & \textbf{66} & \textbf{365} & \textbf{32\%} \\}

%% file: figures/build/t4_rows.tex
\newcommand{\tfourrows}{%
\texttt{claude-fable-5-1} & 36 & 15 & 76.29 & 72.5 & 11 & 210.79 \\
\texttt{gpt-6-astra} & 36 & 16 & 74.48 & 63.25 & 10 & 199.73 \\
\texttt{kimi-k-3} & 38 & 17 & 60.11 & 62.25 & 14 & 191.36 \\
\texttt{grok-4-7} & 35 & 18 & 65.03 & 58.75 & 11 & 187.78 \\
\texttt{gemini-3-1-pro} & 37 & 13 & 44.25 & 44.75 & 10 & 149.00 \\}

%% file: figures/build/t6_rows.tex
\newcommand{\tsixrows}{%
\texttt{claude-fable-5-1} & 48 & 45 & 9 & 6 & 76.29 & 75.97 \\
\texttt{gpt-6-astra} & 45 & 37 & 16 & 8 & 74.48 & 70.52 \\
\texttt{kimi-k-3} & 41 & 40 & 6 & 5 & 60.11 & 62.90 \\
\texttt{grok-4-7} & 35 & 27 & 9 & 1 & 65.03 & 59.52 \\
\texttt{gemini-3-1-pro} & 25 & 29 & 2 & 6 & 44.25 & 53.14 \\
\midrule
\textbf{all} & \textbf{194} & \textbf{178} & \textbf{42} & \textbf{26} & & \\}

%% file: figures/build/t7_rows.tex
\newcommand{\tsevenrows}{%
\texttt{summit\_blizzard} & disabled & 1.67 & \textcolor{noc}{1.67} & \textcolor{noc}{1.67} & \textcolor{noc}{1.67} \\
\texttt{yellow\_white\_flowers} & deleted & 1.67 & \textbf{0} & \textbf{0} & \textbf{0} \\
\texttt{cacti\_on\_flats} & dead code & 1.43 & \textcolor{noc}{1.43} & \textbf{0} & \textcolor{noc}{0.71} \\
\texttt{mangrove\_or\_prop\_roots} & dead code & 1.43 & \textcolor{noc}{1.43} & \textbf{0} & \textcolor{noc}{0.71} \\
\texttt{palm\_silhouettes} & dead code & 0.91 & \textcolor{noc}{0.91} & \textbf{0} & \textcolor{noc}{0.46} \\
\midrule
\multicolumn{3}{@{}l}{\textbf{removed items still credited}} & 4 / 5 & 1 / 5 & 4 / 5 \\
\multicolumn{3}{@{}l}{\textbf{points still given} \textcolor{mute}{(of 7.11)}} & 5.44 & 1.67 & 3.55 \\}
\newcommand{\tsevenuntouched}{7 of 32}
\newcommand{\tsevenbiomes}{5}

%% file: figures/build/loss_macros.tex
\newcommand{\lossPctA}{48.0}

\newcommand{\lossPctB}{32.4}

\newcommand{\lossPctC}{8.9}

\newcommand{\lossPctD}{3.9}

\newcommand{\lossPctE}{6.7}

\newcommand{\lossTotal}{378.3}
\newcommand{\hardBiomeA}{jungle}
\newcommand{\hardPctA}{53}
\newcommand{\hardBiomeB}{delta}
\newcommand{\hardPctB}{51}
\newcommand{\hardBiomeC}{swamp}
\newcommand{\hardPctC}{51}
\newcommand{\hardBiomeD}{grove}
\newcommand{\hardPctD}{50}
\newcommand{\easyBiome}{grassland}\newcommand{\easyPct}{11}
\newcommand{\universalN}{4}

%% file: figures/build/wc002.tex
\newcommand{\wcScore}{15}

%% file: figures/build/wc001.tex
\newcommand{\wcaJudged}{15}
\newcommand{\wcaBuggy}{3}
\newcommand{\wcaBugScore}{16}
\newcommand{\wcaSrcScore}{20}
\newcommand{\wcaRows}{%
\texttt{cube\_primitive} & 1 & 1.00 & \yes & 1 & \texttt{\scriptsize const box=new THREE.BoxGeometry(1,1,1);} \\
\texttt{bulk\_placement} & 1 & 1.00 & \yes & 1 & \texttt{\scriptsize function buildInstanced(list,material)\{ const m=\,…} \\
\texttt{discrete\_grid} & 1 & 1.00 & \yes & 1 & \texttt{\scriptsize const idx=(x,z)=>(x-GX0)+(z-GZ0)*GW,} \\
\texttt{stacked\_columns} & 1 & 1.00 & \yes & 1 & \texttt{\scriptsize B(x,bt,g-capT,z,stoneColor(biome[i],x,z)); B(x,g\,…} \\
\texttt{cube\_terrain} & 1 & 1.00 & \yes & 1 & \texttt{\scriptsize B(x,bt,g-capT,z,stoneColor(biome[i],x,z));} \\
\texttt{grid\_aligned} & 1 & 1.00 & \yes & 1 & \texttt{\scriptsize const x=cellX(i),z=cellZ(i),g=ground[i],bt=Math.\,…} \\
\texttt{unit\_voxels} & 1 & 1.00 & \yes & 1 & \texttt{\scriptsize const box=new THREE.BoxGeometry(1,1,1);} \\
\texttt{contained\_water} & 2 & 1.00 & \yes & 2 & \texttt{\scriptsize const waterMesh=buildInstanced(waters,new THREE.\,…} \\
\texttt{water\_physics} & 3 & 1.00 & \yes & 3 & \texttt{\scriptsize if(d<=w)\{ water[i]=water[i]>-999?Math.min(water[\,…} \\
\texttt{water\_bed} & 5 & 1.00 & \yes & 5 & \texttt{\scriptsize ground[i]=Math.round(sb); water[i]=8; type[i]=sb\,…} \\
\texttt{grounded\_props} & 3 & 1.00 & \yes & 3 & \texttt{\scriptsize B(x,g,g+H,z,rnd()<0.5?0x5a3a20:0x6b4426);} \\}

%% file: figures/build/bd_macros.tex
\newcommand{\bdSpreadWideTest}{WC003}
\newcommand{\bdSpreadWide}{32}
\newcommand{\bdSpreadNarrowTest}{WC001}
\newcommand{\bdSpreadNarrow}{8}
\newcommand{\bdSrcMin}{17}
\newcommand{\bdSrcMax}{20}
\newcommand{\bdBugMin}{16}
\newcommand{\bdBugMax}{20}
\newcommand{\bdCovMin}{9}
\newcommand{\bdCovMax}{10}
\newcommand{\bdRuleMin}{4}
\newcommand{\bdRuleMax}{8}
\newcommand{\bdContentsLo}{grove}
\newcommand{\bdContentsLoAvg}{4.7}
\newcommand{\bdContentsHi}{grassland}
\newcommand{\bdContentsHiAvg}{9.0}
\newcommand{\bdPhysicsLo}{jungle}
\newcommand{\bdPhysicsLoAvg}{4.3}
\newcommand{\bdPhysicsHi}{grassland}
\newcommand{\bdPhysicsHiAvg}{8.7}
\newcommand{\bdLookPts}{65}
\newcommand{\bdMotionPts}{35}
\newcommand{\bdLookPct}{70}
\newcommand{\bdMotionPct}{47}
\newcommand{\bdWcItems}{10}
\newcommand{\bdWcFullN}{2}
\newcommand{\bdWcZeroN}{0}

\newcommand{\bdWcLowList}{\texttt{moon}, \texttt{dusk\_dawn\_tint}, \texttt{sky\_or\_fog\_day\_cycle}}

%% file: sections/00_abstract.tex
\begin{abstract}
Large language models can now write complete, interactive 3D worlds as code, but
grading those worlds automatically is unreliable. Existing judges take one view of
the output: a vision--language model scores a few rendered snapshots, or a language
model reads the source. On worlds written by five frontier models we find that the
two views disagree on 32\% of required items, mostly code that no frame shows, and
that fixed views miss small close-up contents. We present \WB{}, a benchmark and judge for open-ended,
LLM-generated Three.js worlds. From one prompt describing a floating voxel island
with ten biomes, physics, and day/night and seasonal cycles, the judge
\emph{explores} the running world, controlling its clock, orbiting it, and sending a
navigator agent to frame each biome, and \emph{reads} the code for what it sees.
Neither channel is trusted on its own: a code quote counts only if it is text the
source contains, and visual claims are checked against measured pixels where the
property is measurable. A mutation test, in which we remove features by
construction, shows that code-only judging gives full credit to four of five removed features,
because their code remains in the file. Our judge cuts the points kept on removed
features by a third (5.44 to 3.55 of 7.11), and what it still credits is mostly code
that exists but never runs.
We evaluate five frontier models: Claude Fable 5.1, GPT-6 Astra, Kimi K3,
Grok 4.7 and Gemini 3.1 Pro.
Code, prompt, tests and judge configuration are available at
\url{https://github.com/KrishBakshi/worldbench}.
\end{abstract}

%% file: sections/01_intro.tex
\section{Introduction}
\label{sec:intro}

Asked for a 3D world, a language model no longer returns a description or a single
asset. It returns a program: one HTML file that builds terrain from noise
functions, places forests from lookup tables, runs water and weather as particle
systems, and turns night into day on its own clock. What that world \emph{is} only
exists while the program runs. Grading it means answering two questions at once:
does the code build the thing, and does the running world show it? Neither answer
implies the other. Code can build an animal at coordinates no camera ever sees, or
define a weather system that never switches on. A render can show a feature that no
identifier names, such as a terraced noise threshold that \emph{is} a sandstone
mesa.

Automatic evaluators usually answer one of the two. Render-based benchmarks score
a handful of fixed views with a vision--language model
\citep{he2023t3bench,cai2024gt23dbench,tam2025sceneeval}. They see the result but
not its cause, a still frame cannot show a river flowing, and the judge's verdict is
taken as given. For generated Three.js programs, \citet{lu2026worldcoderbench} found
screenshot-plus-VLM judging essentially uncorrelated with ground truth. Code-based
evaluation reads the source, with a model or with patterns. It sees the cause but not
the result, and pattern matching is easy to game: on our own early keyword test, a
one-line file declaring \texttt{const pine=1,\dots,lava=9} scored as highly as every
real world. WorldCoder-Bench sidesteps both by requiring each program to expose its
state through a fixed interface and checking it against expert-written contracts.
That works for functional tasks. It does not fit an open-ended creative prompt:
``the desert borders the jungle'' or ``autumn looks different from spring'' is not
a state variable anyone can specify in advance, and asking for an interface changes
the task being measured.

On five frontier models' worlds, the two views disagree on 32\% of required items.
The code is there and the frames never show it in 105 of 365 cases, and the frames
show it but the code was not found in 10. Disagreement alone does not say which view
is right, so we test that on features we remove by construction
(Section~\ref{sec:mutation}).

We present \WB{}, a benchmark and judge for open-ended, LLM-written 3D
worlds. One prompt asks for a floating voxel island with ten biomes in stated
relations, physical water, lava and weather, and day/night and seasonal cycles.
Five tests with fixed maxima, 280 points in total, score it item by item. The judge
\emph{explores} the running world (Figure~\ref{fig:arch}): it takes control of the world's own clock,
measures which render is day instead of assuming it, orbits the island for wide
views, and sends a navigator agent in for close views of each biome. It also
\emph{reads} the code for what it is looking at. Neither channel is trusted. A code
quote counts only if it is text the source contains, a navigator's ``found it'' is
re-checked blind, and wherever a property is measurable (brightness, motion,
colour shift), pixels decide instead of the visual judge.

Our contributions:
\begin{itemize}
  \item \textbf{A benchmark} for open-ended, code-generated 3D worlds. The prompt asks
  for nothing on the grader's behalf: one fixed prompt, a 280-point ladder of
  item-level tests, and a bounded, error-only generation protocol
  (Section~\ref{sec:benchmark}).
  \item \textbf{A judge that explores and reads, and trusts neither}, with every
  model-made claim backed by the source text or by measured pixels before it scores
  (Section~\ref{sec:method}).
  \item \textbf{Evidence for each design choice} on five frontier models' worlds: an
  agreement analysis, a close-versus-wide ablation, a mutation test with features
  removed by construction, and an error analysis over four error classes
  (Section~\ref{sec:results}).
\end{itemize}

%% file: sections/02_related.tex
\section{Related work}
\label{sec:related}

\paragraph{Evaluating generated 3D content.}
T\textsuperscript{3}Bench \citep{he2023t3bench} scores text-to-3D methods from
multi-view renders, with a quality metric and an alignment metric in which a
language model judges multi-view captions; it reports that current methods struggle
most with surroundings and multi-object scenes, the kind of content our island
asks for. GT23D-Bench \citep{cai2024gt23dbench} pairs a large annotated asset set
with ten metrics validated by correlation with human judgment. SceneEval
\citep{tam2025sceneeval} evaluates indoor scene synthesis on \term{explicit requirements}
(object counts, attributes, relations) and \term{implicit expectations} (support,
collisions, navigability), using a vision--language model on fixed renders and
geometric checks on the scene's meshes wherever possible. Our split between
required contents (WC003) and physical behaviour (WC004) follows the same
explicit/implicit distinction. What differs is the object being judged: these
benchmarks score assets or scene data, while we score a program whose output exists
only when it runs, and whose camera, clock and seasons we must drive ourselves.

\paragraph{Language models that build 3D scenes.}
A line of systems uses language models to \emph{generate} scenes rather than grade
them. SceneCraft \citep{hu2024scenecraft} writes Blender Python from a scene graph
and refines it with a vision--language model that critiques the render. SceneX
\citep{zhou2024scenex} plans procedural Blender commands over a curated asset
library and demonstrates natural landscapes and seasonal changes, the content
closest to ours. Reason-3D \citep{berdoz2025reason3d} uses large reasoning models for
object retrieval and collision-aware placement, and SceneOrchestra
\citep{he2026sceneorchestra} trains an orchestrator to emit whole tool-call
trajectories instead of rendering after every step. These systems give the model
scene tools and asset libraries. Our models get one prompt and plain Three.js and
must write the procedural generation themselves, repairing from errors only
\citep[in the spirit of][]{chen2023selfdebug}.

\paragraph{Benchmarks of LLM-written interactive worlds.}
WorldCoder-Bench \citep{lu2026worldcoderbench} is the closest work: 2{,}026 tasks
whose output, like ours, is one self-contained Three.js page. It runs each program
in a headless browser, requires it to expose a standard state object, applies
scripted actions, and checks state changes against hidden contracts written by
experts. It also reports that screenshot-based judging fails on these programs.
We share the medium and the scepticism about visual judges, but take the opposite
task shape: one open-ended world instead of thousands of narrow functional tasks,
with nothing exposed for the grader. Where it replaces visual judging with
contracts, we keep visual judging and verify it, against the source text and
against pixels. Table~\ref{tab:related} compares the benchmarks side by side.

\begin{table}[!tb]
  \centering\small
  \input{figures/tab_related_body.tex}
  \caption{Where \WB{} sits among the evaluation benchmarks it cites (generators are
  discussed in the text). \partly: partly.}
  \label{tab:related}
\end{table}

\paragraph{Models as judges.}
Strong language models agree with human preferences often enough to serve as
judges \citep{zheng2023judging}, but they carry biases, including a preference for
their own family's outputs \citep{panickssery2024self}. Most judge pipelines respond
by calibrating the judge against people. Our worlds are open-ended and synthetic,
so there is no reference answer to annotate. We instead check each judgement
against evidence the model under test cannot fake, and validate the judge by
mutation \citep{jia2011mutation}: features removed on purpose, where the correct
verdict is known by construction.

%% file: figures/tab_related_body.tex
\begin{tabularx}{\textwidth}{@{}l >{\hsize=1.5\hsize}Y >{\hsize=.8\hsize}Z >{\hsize=.7\hsize}Z Z Z Z@{}}
\toprule
 & \textbf{what is judged} & \textbf{looks at renders} & \textbf{reads code} &
   \textbf{explores the world} & \textbf{each verdict checked} & \textbf{nothing added to output} \\
\midrule
T\textsuperscript{3}Bench & 3D asset & \yes\ fixed views & -- & \no & \no & \yes \\
GT23D-Bench & 3D asset & \yes\ fixed views & -- & \no & \no & \yes \\
SceneEval & indoor scene (objects, meshes) & \yes\ fixed views & -- & \no & \partly\ geometry where possible & \yes \\
WorldCoder-Bench & Three.js program, functional task & \no & \no & \partly\ scripted actions & \yes\ state contracts & \no\ must expose a state object \\
\midrule
\textbf{worldbench} & Three.js program, open-ended world & \yes & \yes & \yes\ time, seasons, per-biome navigator & \yes\ source text + pixels & \yes \\
\bottomrule
\end{tabularx}

%% file: sections/03_benchmark.tex
\section{The benchmark}
\label{sec:benchmark}

\begin{figure}[!tb]
  \centering
  \includegraphics[width=\textwidth]{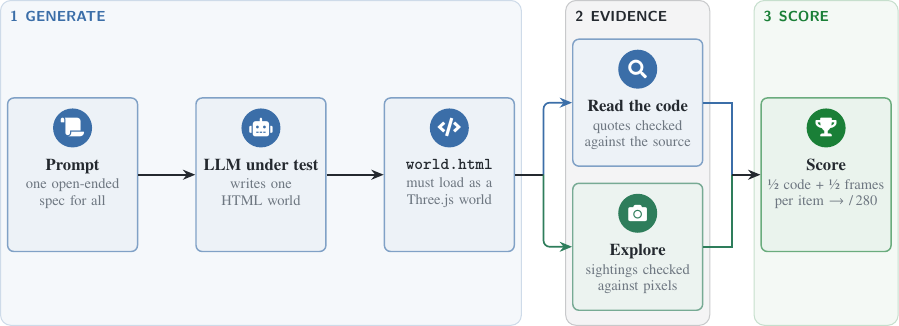}
  \caption{The evaluation flow. The model under test turns one fixed prompt into a
  single \texttt{world.html}, which must load as a Three.js world. Evidence is then
  gathered two ways, by reading the code and by exploring the running world, and each
  kind is checked (quotes against the source, sightings against pixels).}
  \label{fig:flow}
\end{figure}

Figure~\ref{fig:flow} shows the whole flow, from prompt to score. This section
covers the prompt, how worlds are produced, and how they are scored;
Section~\ref{sec:method} covers how each item's evidence is gathered and checked.

\paragraph{One open-ended prompt.}
Every model receives the same prompt (about 2{,}000 words, reproduced in
Appendix~\ref{app:prompt}). It asks for a single, self-contained HTML file that
renders, with Three.js, a floating voxel island in a black void: ten biomes (snow
mountains, snowy conifer forest, highlands, jungle, swamp, flowering grove,
grassland plateau, coastal delta, desert, volcano) placed in stated relations to
one another (the forest sits in front of the mountains, the swamp on the jungle's
far side, the volcano away from the desert); water, lava and particles that behave
physically; a day/night cycle and four seasons; and a small on-screen legend. The
prompt fixes relationships, not coordinates, and says nothing about how the world
will be graded. There is no ground-truth world to compare against. The prompt asks for nothing
on the grader's behalf. Grading runs a copy with one clock patch
(Section~\ref{sec:method}).

\paragraph{Generation.}
We evaluate five models, each queried through its developer's official API:
Claude Fable 5.1 (Anthropic), GPT-6 Astra (OpenAI), Kimi K3 (Moonshot AI),
Grok 4.7 (xAI) and Gemini 3.1 Pro (Google DeepMind); Table~\ref{tab:models} lists
the generation settings. Each world is one completion, sampled at temperature 1.0, continued if the response is truncated. Before
grading, a bounded \term{repair loop} removes crashes that would otherwise hide the whole
world: the file is parsed (\texttt{node --check}) and loaded in a headless browser,
and any parse error or console error is sent back to the model, which may edit the
file with small verified replacements, for at most five rounds. The model is told
this budget up front. Repairs see only errors, never renders or scores, so
generation never sees the grader's view.

\paragraph{The ladder.}
A world is graded by five \term{tests} with fixed maxima, 280 points in total
(Table~\ref{tab:ladder}), so totals compare across models. Each test scores many
small, checkable \term{items} rather than one holistic grade, and every item is
decided by the judge described next. Throughout the paper, worked examples use one
\term{reference world}, \texttt{claude-fable-5-1}, the highest-scoring world.

\begin{figure}[!htb]
  \centering
  \includegraphics[width=\textwidth]{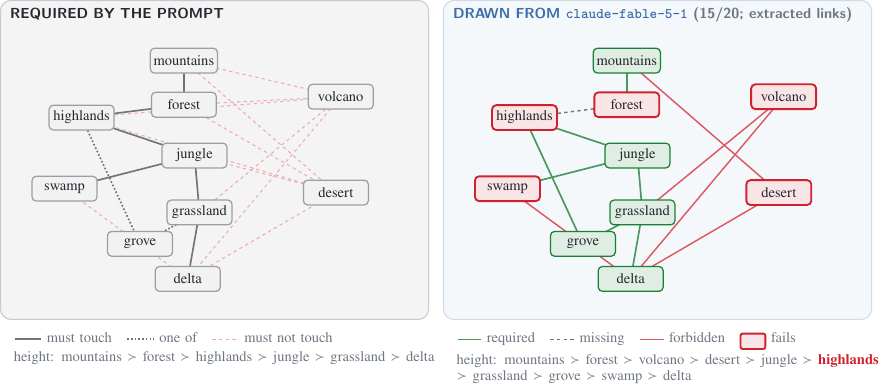}
  \caption{WC002, placement, on the reference world. \emph{Left}: the relations the
  prompt requires, as the grader encodes them. \emph{Right}: the links the judge model
  read from \texttt{claude-fable-5-1}'s code, restricted to the pairs the rules
  mention. It is not a map of the island: node positions are the same fixed layout on
  both sides. Each biome is coloured by its WC002 verdict; the elevation order read
  from the code is printed below, with the one violation in red. The forest and
  highlands fail on a missing link, the highlands also sit below the jungle, and the
  swamp, desert and volcano each touch a biome they must not. Against this world's
  actual biome map, two of these 19 pairs are misread: the forest and highlands do
  touch, and the desert does not touch the mountains (it still fails on its real
  border with the delta).}
  \label{fig:wc002}
\end{figure}

WC002 is the one test of spatial reasoning: the prompt describes the island only
through relations, and the world must turn them into a map. A model reads the whole
source, finds where the layout decides each cell's biome (region centres, boxes,
bands, coast rules) and reports which biomes touch and how high each sits. It may
list two biomes as touching only when the positions it found put them side by side,
quoting those numbers. Fixed rules written from the prompt then score that report
(Figure~\ref{fig:wc002}). The reference world scores \wcScore{} of 20 here. The reading is a
model's, so Section~\ref{sec:limitations} checks it against the actual biome maps of
four worlds.

%% file: figures/tab_ladder_body.tex
\begin{tabularx}{\textwidth}{@{}l >{\hsize=.86\hsize}Y >{\hsize=1.14\hsize}Y r@{}}
\toprule
\textbf{test} & \textbf{what it measures} & \textbf{evidence} & \textbf{max} \\
\midrule
WC001 & it is a voxel island: cubes on a grid, contained water, floating over a void
      & code judge (probability + verbatim quote per item) \emph{and} a visual bug hunt over the frames & 40 \\
WC002 & all 10 biomes present and placed as asked (neighbours, elevation order)
      & a biome counts only if its layout code is really in the source; a model reads neighbours and heights from the whole source; fixed rules score that reading & 20 \\
WC003 & each biome's required contents (peaks, cacti, mangroves, \dots) and nothing that belongs elsewhere
      & per item: verified code quote + navigated biome frame; disagreements re-searched in code & 100 \\
WC004 & physics and motion: water in beds, things fall, rivers flow, particles drift
      & per item: verified code quote + frames; motion counts only if a burst's pixels changed & 100 \\
WC005 & day/night and seasons: sun orbit, dimming, tint, season cycle
      & code probe; change measured from pixels (brightness, frame difference, chromaticity) & 20 \\
\midrule
 & & \textbf{total} & \textbf{280} \\
\bottomrule
\end{tabularx}

%% file: sections/04_method.tex
\section{The judge: explore and read, trust neither}
\label{sec:method}

Figure~\ref{fig:arch} shows the idea. \emph{Reading} looks at the cause: the program
as written. \emph{Looking} looks at the result: the program as it runs. An item
scores only for the evidence that survives its check. Several
models fill the judge's roles; they and their settings are listed in
Appendix~\ref{app:judges}. Each check below exists because a simpler judge failed on
real worlds; Appendix~\ref{app:failures} lists those judges and how each failed.

\subsection{Looking: exploring the running world}
\label{sec:looking}

The saved \texttt{world.html} is not edited. Grading runs a copy whose clock is
patched, as in Stage I, in a headless browser. An init script seeds
\texttt{Math.random}, so a reload builds the same island, and reaches the scene and
camera through Three.js's own developer-tools hook.

Looking runs as a four-stage pipeline: the world's time is brought under control,
views are taken, motion is recorded, and every close view is checked blind before
it can be judged.

\begin{figure}[!htb]
  \centering
  \includegraphics[width=\textwidth]{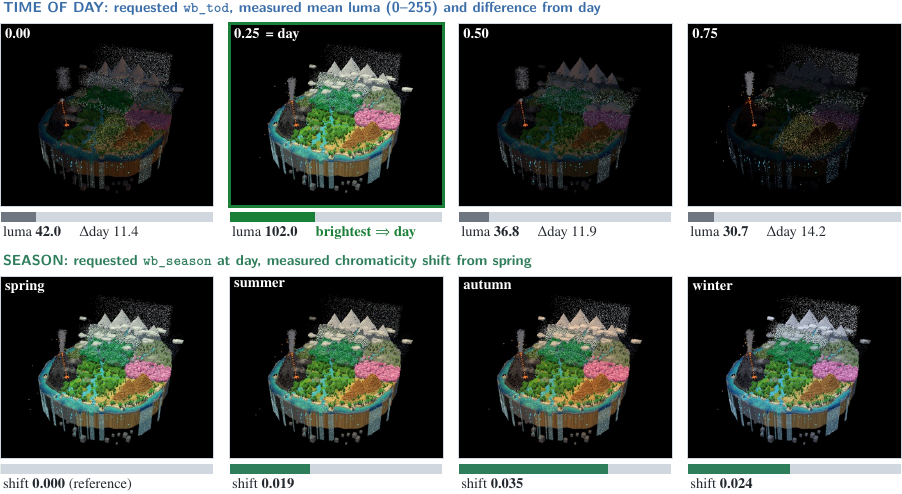}
  \caption{Controlling time, reference world. \emph{Top}: the world's own clock,
  patched to the requested time of day; the brightest render (here requested time
  0.25, not 0.50) becomes day for every later view, and each other render's frame
  difference from day ($\Delta$day) is recorded. \emph{Bottom}: the four seasons at
  day. The visual judge called these frames identical; the measured chromaticity
  shift from spring is non-zero for every season, and it is the measurement that
  scores the item.}
  \label{fig:clock}
\end{figure}

\paragraph{Stage I: Controlling time.}
Every world keeps its own clock, written its own way. A language model patches that
clock so it obeys an externally set time of day and season. The patch is tightly
constrained: at most six edits, each under 600 added characters, each required to
read the external time, so a patcher cannot add lighting the model never wrote.
The patch's mapping from requested time to noon is not trusted either. The world is
rendered at four requested times and the brightest render, by mean luma, is taken
as day (Figure~\ref{fig:clock}, top). In the reference world, noon falls at a quarter
of the cycle, where a fixed ``half-way is noon'' rule would have chosen dusk. The
four seasons are then rendered at day (Figure~\ref{fig:clock}, bottom). Change over
time is measured on these frames, not judged: mean luma for night, frame difference
against day, and chromaticity shift for twilight tint and seasons. In the reference
world the visual judge described the four season frames as ``\wbSeasonSeen''; their
measured chromaticity shift from spring is \wbChromaB, \wbChromaC{} and \wbChromaD,
and the measurement decides.

\begin{figure}[!htb]
  \centering
  \includegraphics[width=\textwidth]{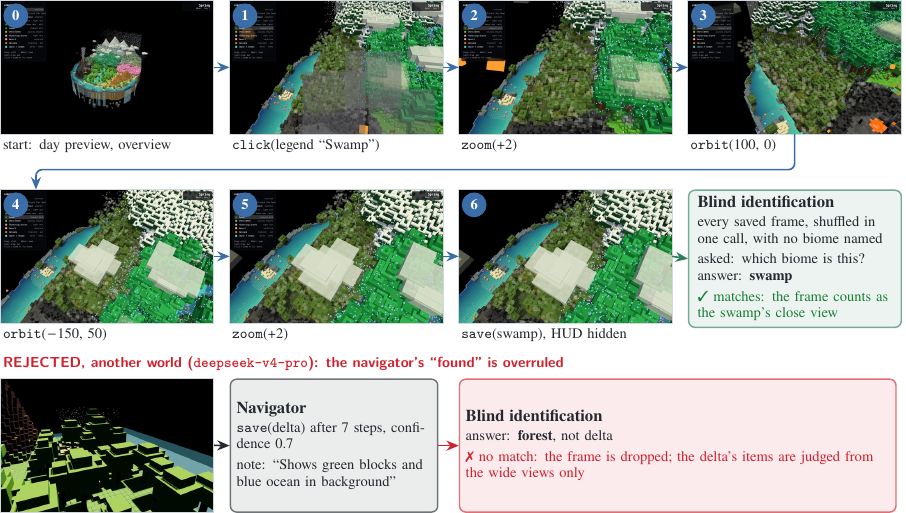}
  \caption{One navigator episode, reference world, swamp. Each frame is what the
  agent saw after the action below it; frames are re-rendered from the recorded
  actions with no model call (seeded world, pinned clock, so only animation phase
  differs from the original run). The saved frame (6) counts only after blind
  identification agrees. \emph{Bottom}: a rejection from another graded world outside
  the five reported here; the navigator saved the delta with confidence
  \wbRejConf, and the blind answer was \wbRejIdent.}
  \label{fig:nav}
\end{figure}

\paragraph{Stage II: Wide and close views.}
From day, the judge takes \term{wide views} (an overview, four orbits, and the four
seasons) and \term{close views} of each biome. Close views come from a
\term{navigator agent}, a tool-calling loop of observe $\rightarrow$ act $\rightarrow$
observe. Each turn it receives the page's accessibility tree and its latest frame,
and calls exactly one tool: \texttt{click} (a legend entry the world itself
provides), \texttt{orbit}, \texttt{pan}, \texttt{zoom}, \texttt{save} or
\texttt{give\_up}, within an eight-step budget. Figure~\ref{fig:nav} shows the
reference world's longest episode: the navigator clicks the swamp's legend entry,
zooms in, orbits away and back to bring the water channels into frame, zooms again
and saves (\wbNavSteps{} steps). Each turn keeps only the newest frame in context,
so a long episode does not grow the prompt.

\begin{figure}[!htb]
  \centering
  \includegraphics[width=\textwidth]{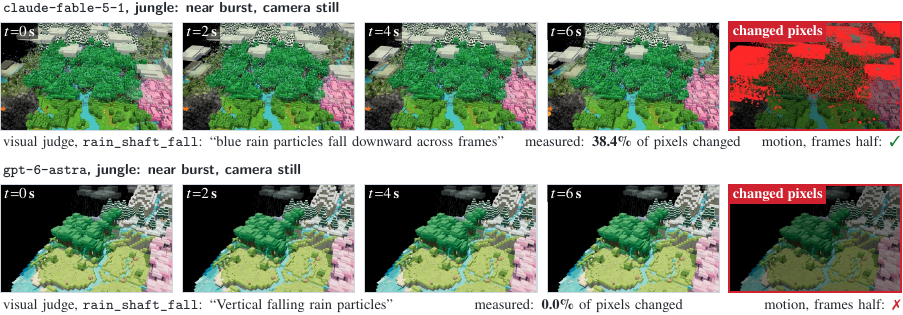}
  \caption{Still-camera motion bursts for one WC004 item, \texttt{\wbBurstItem}, in
  the jungles of \texttt{\wbBurstModelA} (left) and \texttt{\wbBurstModelB} (right). Red in the overlay marks pixels that changed anywhere in the
  burst. The visual judge reports falling rain in both; only the burst whose pixels
  changed earns the frames half of the motion credit.}
  \label{fig:burst}
\end{figure}

\paragraph{Stage III: Motion bursts.}
A single frame cannot show motion, and a visual judge asked about motion across
frames will report it from appearance alone. For each saved close view the
judge therefore takes a \term{still-camera burst}: four frames \wbBurstGap{}\,s apart
with the camera fixed, then one zoom out and three more, and an overlay that paints
every changed pixel. A visual ``it moves'' counts only if the burst's pixels
changed. Figure~\ref{fig:burst} shows the same motion item in two worlds. In one,
rain falls and \wbBurstFracA\% of pixels change. In the other, the visual judge
again reported ``\wbBurstSeenB'', but \wbBurstFracB\% of pixels changed, so the
claim is refused. A typical world yields about a hundred frames in total.

\paragraph{Stage IV: Blind identification.}
The navigator's claim that it found a biome is not trusted: in one world it saved
the volcano as the swamp with full confidence. Every saved frame is re-taken with
the page's own interface hidden, so the clicked legend label cannot give the answer,
and then \term{identified blind}: all of a world's saved frames go to one shuffled
call that is never told which biome a frame was meant to show. A frame counts as a
biome's close view only if the answer matches (Figure~\ref{fig:nav}, middle row);
otherwise it is dropped and that biome is judged from the wide views alone
(Figure~\ref{fig:nav}, bottom row). Appendix Figure~\ref{fig:sheet} shows every
close view of the reference world.

\subsection{Reading: the code as evidence}
\label{sec:reading}

Reading also runs in stages: quotes are proposed, verified against the source,
repaired when badly copied, and searched again where the frames disagree.

\begin{figure}[!tb]
  \centering
  \includegraphics[width=\textwidth]{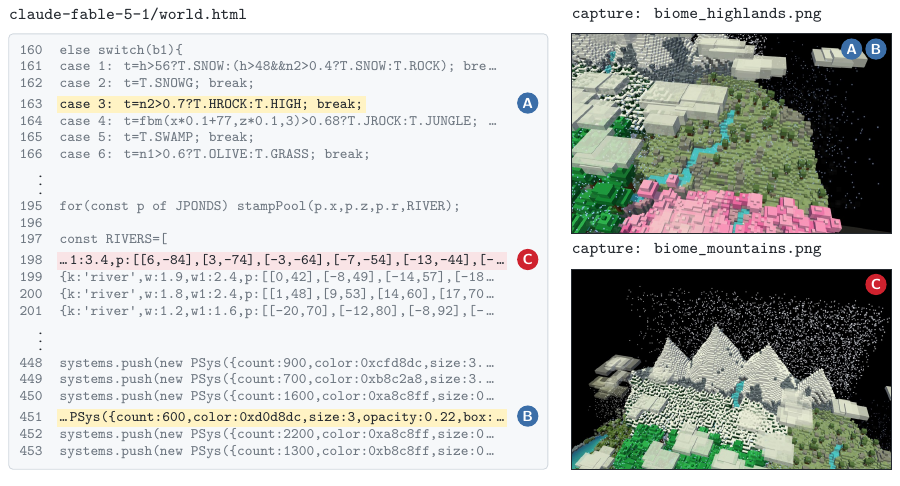}\\[6pt]  
  \small\input{figures/fig_verified_body.tex}
  \caption{Three items, verified two ways (reference world; A and B in the highlands,
  C in the mountains). \textbf{A}: the quote is source line~\wbLineA{} verbatim and the
  frame shows stone, so both halves are earned. \textbf{B}: the quote is real
  (line~\wbLineB), but no frame shows the mist, so only the code half is earned (error class E2, \emph{unobserved}; Section~\ref{sec:errors}).
  \textbf{C}: the frame shows the meltwater stream, but the code probe found nothing in
  the mountains and the recheck agent did not credit it either: the stream comes from a
  river table placed by coordinates (line~\wbLineC), not from code that names the
  biome. Only the frames half is earned.}
  \label{fig:verified}
\end{figure}

\paragraph{Stage I: Quoting.}
For each biome, a \term{code probe} lists the test's items and must quote, for each one
it claims, the source lines that build it.

\paragraph{Stage II: Verifying.}
A quote counts only if it is text the
source contains: matching ignores whitespace and is done line by line, at least
80\% of a quote's lines must be found, and a line whose ending was garbled in
copying still matches if its first 85\% is verbatim. Invented code does not pass.
Before this check, a model was credited for a plausible
\texttt{scene.add(new THREE.Mesh(palmGeo,\dots))} that appeared nowhere in the file.
A quote must also come from that biome's own code, and one quote cannot satisfy
two items.

\paragraph{Stage III: Repairing.}
Probes copy imperfectly, joining separate lines or abbreviating with an
ellipsis. Rather than loosen the check, a \term{quote repair} step lets a model choose
\emph{line numbers} from a numbered window of the file, and rebuilds the quote from
those lines, so it can select code but never write it.

\paragraph{Stage IV: Rechecking.}
When the frames show an item but the probe found no code for it, an \term{evidence agent}
searches the source with read-only tools (\texttt{grep}, line reads and a small set
of text-only shell commands) over a sandboxed copy of the file, within ten tool
calls. It is told that a sighting is a lead, not proof, and that inferring a
feature from a neighbouring one (``there are pools, so there must be banks'') is
not evidence. Only these disagreements are searched again. This is the one stage
that waits for Looking, since it needs the frames.

\subsection{Combining the two}
\label{sec:combine}

An item's points are split evenly between the two channels
(Figure~\ref{fig:verified}). The \term{code half} is earned by a quote that passes the
checks above. The \term{frames half} is earned when the visual judge reports the item in
a blind-confirmed frame of that biome, or, where the property is measurable, when
the pixels confirm it (Stages I and III). Items that must \emph{not} appear in a biome (a palm in the desert, snow on the
volcano) cost points when either channel finds them: a verified quote that builds one
in the biome, or a sighting in its frames.

\paragraph{Scoring.}
Within a test, each biome is worth 10 points, divided among its required items by
weight $w_i$, so item $i$ is worth
\begin{equation} p_i = 10\,\frac{w_i}{\sum_j w_j}. \label{eq:points} \end{equation}
With $c_i\in\{0,1\}$ for a passed code check (Stage~II of Reading) and $v_i\in\{0,1\}$
for a sighting in a blind-confirmed frame (Stage~IV of Looking), the item earns
\begin{equation} g_i = p_i\left(\tfrac12 c_i + \tfrac12 v_i\right). \label{eq:credit} \end{equation}
An item that must also move (WC004) splits its points between look and motion, and its
motion half is split again between code and the burst:
\begin{equation} g_i = p_i\left[\tfrac12\left(\tfrac12 c_i+\tfrac12 v_i\right)
  + \tfrac12\left(\tfrac12 c^{\mathrm{mot}}_i + \tfrac12 m_i\right)\right], \label{eq:motion} \end{equation}
where $m_i=1$ only if the visual judge reports motion on the expected axis \emph{and}
the burst's changed-pixel share $f_i \ge 0.001$. Each forbidden item $l$ that is
coded or seen subtracts its share, and a biome cannot go below zero:
\begin{equation} S_b = \max\Bigl(0,\ \sum_i g_i - \sum_l p_l\,[c_l \lor v_l]\Bigr),
  \qquad \mathrm{WC003} = \sum_{b=1}^{10} S_b. \label{eq:biome} \end{equation}
Item~B in Figure~\ref{fig:verified}, for example, has $c=1$ and $v=0$, and earns
$\tfrac12\times\wbPointsB = \wbEarnedB$. Appendix~\ref{app:scoring} gives the
formulas for the other tests and the pixel measures.

Because the halves are independent, code that never renders earns at most half, and
so does a render the probe could not explain; Section~\ref{sec:results} measures how
often each happens.

%% file: figures/fig_verified_body.tex
\setlength{\tabcolsep}{3.5pt}%
\begin{tabularx}{\textwidth}{@{}c >{\hsize=.85\hsize}Y >{\hsize=1.3\hsize\ttfamily\scriptsize}Y c >{\hsize=.85\hsize}Y c r@{}}
\toprule
 & \textbf{item} & \textrm{\normalsize\textbf{probe's code quote}} & \textbf{in src} &
   \textbf{frame judge says} & \textbf{halves} & \textbf{pts} \\
\midrule
\mA & \wbLabelA & \wbQuoteA & \wbCodeA & \wbSeenA & \pbar{\wbCodeFillA}{\wbVisFillA} & \wbEarnedA\,/\,\wbPointsA \\[2pt]
\mB & \wbLabelB & \wbQuoteB & \wbCodeB & \wbSeenB & \pbar{\wbCodeFillB}{\wbVisFillB} & \wbEarnedB\,/\,\wbPointsB \\[2pt]
\mC & \wbLabelC & \wbQuoteC & \wbCodeC & \wbSeenC & \pbar{\wbCodeFillC}{\wbVisFillC} & \wbEarnedC\,/\,\wbPointsC \\
\bottomrule
\end{tabularx}

%% file: sections/05_results.tex
\section{Results}
\label{sec:results}

\paragraph{Setup.}
We grade one world from each of the five models of Section~\ref{sec:benchmark}, all
with the same judge configuration. We ask three questions of the judge before reading the leaderboard: do its two
channels disagree, does exploring matter, and does it notice when a feature is
really gone?

\subsection{Code and frames disagree on a third of items}
\label{sec:agreement}

WC003 checks 73 required items per world, 365 in total. For each, the judge has a
code verdict and a frames verdict (Table~\ref{tab:agreement}). They agree on 250
items and disagree on 115 (32\%), between 26\% and 41\% per world. Disagreement is
lopsided: in 105 items the code is found but no frame shows the item, against 10
in the other direction. A judge that only read code would credit 289 items (79\%);
one that only looked would credit 194 (53\%). ``Not seen'' means not seen in our
frames, which is either a real absence or an item our capture did not reach;
Section~\ref{sec:mutation} separates the two on items we removed ourselves, and
Section~\ref{sec:errors} names the two disagreement directions as error classes E2
(\term{unobserved}) and E3 (\term{unattributed}).

\begin{table}[!htb]
  \centering\small
  \input{figures/tab_agreement_body.tex}
  \caption{Code and frames verdicts on WC003's required items, per world. ``Not
  seen'' means not seen in our frames.}
  \label{tab:agreement}
\end{table}

\subsection{Close and wide views see different things}
\label{sec:closewide}

To test whether exploring matters, we re-ran WC003's visual judge on the same saved
frames with the navigator's close views removed, leaving the overview and four
fixed orbits, which is what a \term{snapshot judge} has. Prompt, items and judge are
unchanged (Table~\ref{tab:snapshot}). Close views see more required items (194
against 178), but not by a wide margin, and the two sets differ: 42 items are seen
only up close and 26 only in the wide views. The close-only items are small
contents (animals account for 12 of the 42, followed by flowers, tall grass, vines
and rain shafts). The wide-only items are biome-scale layout and weather (swamp
channels, river mouths, islands, flat basins, mist, ash). The code probe
independently supports 81\% of the extra sightings on both sides, so neither set
is mostly hallucinated. Totals barely move (320.2 against 322.1 WC003 points over
five worlds), but individual worlds do: \texttt{grok-4-7} loses 5.5 points without
close views and \texttt{gemini-3-1-pro} gains 8.9, which swaps \texttt{kimi-k-3}
and \texttt{grok-4-7} in the WC003 ranking. Exploration matters for coverage, not
because any one camera is better. Two caveats apply: the close condition gave the
judge two images per biome and the wide one five, and the navigator confirmed
only 5 of 10 biomes in \texttt{grok-4-7}'s world.

\begin{table}[!htb]
  \centering\small
  \input{figures/tab_snapshot_body.tex}
  \caption{Close (navigated) vs.\ wide (fixed snapshot) views: same judge, items and
  saved frames; only the images given to the judge differ.}
  \label{tab:snapshot}
\end{table}

\subsection{Removing features on purpose}
\label{sec:mutation}

Since there is no reference answer to annotate, we create \term{ground truth by
construction}, as in mutation testing \citep{jia2011mutation}. In
a copy of \texttt{claude-fable-5-1}'s world we remove features, then re-run capture
and WC003 on the changed biomes (Table~\ref{tab:mutation}). One feature is deleted
outright. Three are made \emph{dead code}: the function that builds them stays in
the file but is never called. A fourth is \term{disabled in place} (\texttt{if(false)}).
The world is otherwise identical, including its random stream. The correct verdict
for every removed item is absent.

\begin{table}[!htb]
  \centering\small
  \input{figures/tab_mutation_body.tex}
  \caption{Mutation test on a copy of \texttt{claude-fable-5-1}'s world: points each
  judge still gives to items we removed. The correct value for every row is 0.}
  \label{tab:mutation}
\end{table}

A code-only judge credits all four features whose code is still in the file, at
full value, quoting real lines such as the kept \texttt{function cactus(\dots)}
definition. Our frames half sees none of the cacti, mangroves or palms and cuts
each to half credit: removed items keep 3.55 of their 7.11 points under our judge,
against 5.44 under code-only judging. One removal slips through entirely: the
disabled blizzard, where the probe quoted the switched-off line and the visual
judge read drifting snowfall near the peaks as a blizzard. A snapshot-only judge
is strictest here (1.67 points kept), but Section~\ref{sec:closewide} shows it also
misses real close-up items. Two findings bound these numbers. First, our code half
still accepts code that never runs, so dead code earns half credit
(Section~\ref{sec:limitations}). Second, single-run
item verdicts are noisy: 7 of 32 items we did not touch changed verdict between the
original run and this one. A sixth removal (volcanic ash) is excluded because the
item also accepts eruption particles, and the volcano's lava bombs, which we did
not remove, still provide them; the judge was right to keep crediting it.

\subsection{Leaderboard}
\label{sec:leaderboard}

\begin{figure}[!htb]
  \centering
  \includegraphics[width=\textwidth]{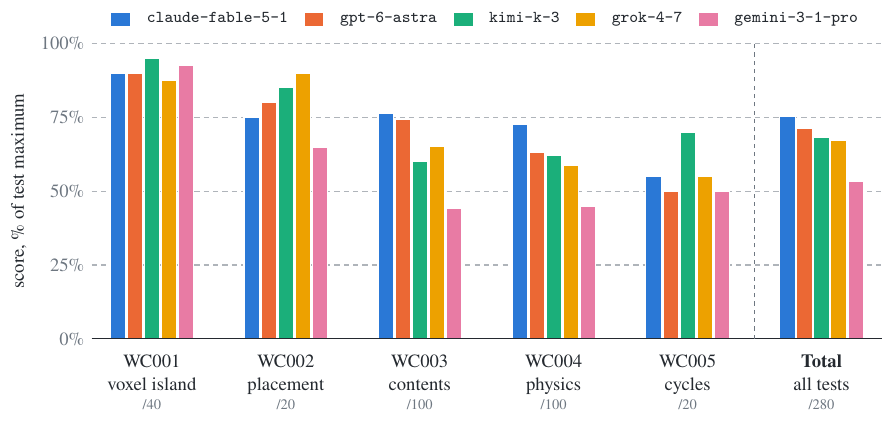}
  \caption{Per-test scores as a percent of each test's maximum (maxima differ, 20 to
  100 points), and the total out of 280. Exact values are in Table~\ref{tab:leaderboard}.}
  \label{fig:scores}
\end{figure}

Table~\ref{tab:leaderboard} gives the totals and Figure~\ref{fig:scores} the
per-test profile. \texttt{claude-fable-5-1} leads with
210.79 of 280, followed by \texttt{gpt-6-astra} (199.73), \texttt{kimi-k-3}
(191.36), \texttt{grok-4-7} (187.78) and \texttt{gemini-3-1-pro} (149.00). The order
is not uniform across tests. \texttt{claude-fable-5-1} leads on contents (76.29) and
physics (72.5). \texttt{kimi-k-3} is ahead on the voxel island (38/40) and on cycles
(14/20), and \texttt{grok-4-7} on placement (18/20); those gaps are a few points, and
the placement column is a model's reading of the source (Section~\ref{sec:limitations}).
We read gaps of a few points between neighbours as ties, in totals as well, such as
\texttt{kimi-k-3} and \texttt{grok-4-7}.

\begin{table}[!htb]
  \centering\small
  \input{figures/tab_leaderboard_body.tex}
  \caption{Leaderboard, ordered by total. Gaps of a few points between neighbours are within judge noise (Section~\ref{sec:mutation}).}
  \label{tab:leaderboard}
\end{table}

\subsection{Error analysis}
\label{sec:errors}

We classify every point lost on WC003 and WC004 (\lossTotal{} points over the five
worlds) by the state of its item's two channels. The code channel is positive when
a verified quote exists (Section~\ref{sec:reading}); the frames channel is positive
when a blind-confirmed frame shows the item (Section~\ref{sec:looking}). This gives
four error classes (Table~\ref{tab:errclass}), used with these names throughout.

\begin{table}[!htb]
  \centering\small
  \begin{tabularx}{\textwidth}{@{}l c c Y r@{}}
  \toprule
  \textbf{error class} & \textbf{code} & \textbf{frames} & \textbf{meaning} & \textbf{share of lost points} \\
  \midrule
  E1 \term{unimplemented} & \no & \no & no verified quote and no confirmed frame & \lossPctA\% \\
  E2 \term{unobserved} & \yes & \no & the code is in the source; no capture shows the item & \lossPctB\% \\
  E3 \term{unattributed} & \no & \yes & a frame shows the item; no quote verifies against the source & \lossPctC\% \\
  E4 \term{property violation} & \yes & \yes & look confirmed in both; the item's motion is not verified (the quoted motion code is not a per-frame update, or the burst's pixels do not change) & \lossPctD\% \\
  \midrule
  \multicolumn{2}{@{}l}{penalties and gates} & & not an error class: a wrong-biome feature coded or seen, the delta zeroed for an ocean over the void (WC003), or the island gate (WC004) & \lossPctE\% \\
  \bottomrule
  \end{tabularx}
  \caption{Error classes for a lost item, by which channel's evidence is missing.
  Shares are of all \lossTotal{} points lost on WC003 + WC004 across five worlds, which is
  1000 minus their WC003 and WC004 scores.}
  \label{tab:errclass}
\end{table}

E1 dominates (\lossPctA\%). E2 is the second largest (\lossPctB\%), and it is
exactly the set a code-only judge would have credited, so a third of all losses lie
in the gap this benchmark measures. E3 (\lossPctC\%) is the reverse gap, closed in
part by the evidence-agent recheck; E4 is rare (\lossPctD\%), and the last \lossPctE\% are penalties and gates, which no channel explains. The class definitions
are operational, not causal: E1 is an upper bound on true absence, because it is
the code probe's verdict, and E2 includes items that exist but that capture did not
reach (Section~\ref{sec:limitations}).

\begin{figure}[!htb]
  \centering
  \includegraphics[width=\textwidth]{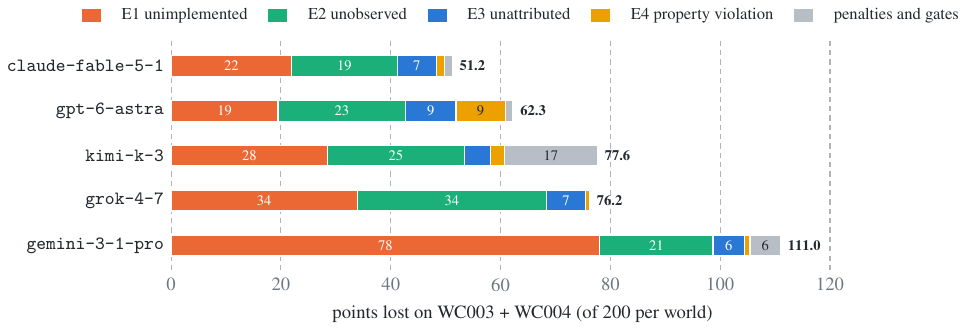}
  \caption{Lost WC003 + WC004 points per model, by error class
  (Table~\ref{tab:errclass}); each bar is 200 minus the world's WC003 and WC004 scores.
  Green (E2, unobserved) is the code-only judge's blind spot: code that exists but that
  no capture shows. Grey is penalties and gates, mostly \texttt{kimi-k-3}'s delta, zeroed
  for an ocean over the void.}
  \label{fig:losses}
\end{figure}

\paragraph{By model.} The class mix differs per world (Figure~\ref{fig:losses}).
\texttt{gemini-3-1-pro}'s losses are mostly E1. \texttt{grok-4-7} has the most E2
points, partly a capture effect: the navigator confirmed only five of its ten biomes
(Section~\ref{sec:closewide}). \texttt{gpt-6-astra} has the most E4 points.

\begin{figure}[!htb]
  \centering
  \includegraphics[width=\textwidth]{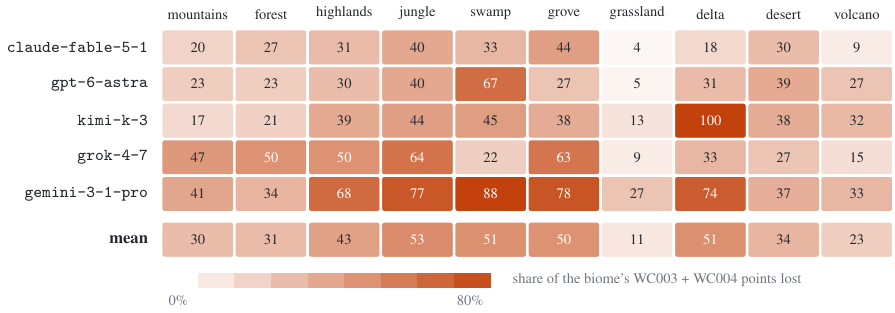}
  \caption{Share of each biome's WC003 + WC004 points lost, per model and on average.
  The jungle is among the four hardest biomes for every model, and the open grassland is
  the easiest for every model.}
  \label{fig:heatmap}
\end{figure}

\paragraph{By biome.} Wet and detailed regions lose the most points
(Figure~\ref{fig:heatmap}). The jungle is among the four hardest biomes for every model,
and the grassland is the easiest for every model. On average \hardBiomeA{} loses \hardPctA\% of its points,
\hardBiomeB{} \hardPctB\%, \hardBiomeC{} \hardPctC\% and \hardBiomeD{} \hardPctD\%,
against \easyPct\% for the open \easyBiome. \universalN{} items are E1 for all five
models, all small life (forest-floor cover and ferns, the grove's pollinators and
bees).

%% file: figures/tab_agreement_body.tex
\begin{tabular*}{\textwidth}{@{\extracolsep{\fill}}l rrrr r r@{}}
\toprule
 & \multicolumn{2}{c}{\textbf{code: found}} & \multicolumn{2}{c}{\textbf{code: not found}} & & \\
\cmidrule(lr){2-3}\cmidrule(lr){4-5}
\textbf{world} & frames: seen & not seen & seen & not seen & \textbf{items} & \textbf{disagree} \\
\midrule
\tthreerows
\bottomrule
\end{tabular*}

%% file: figures/tab_snapshot_body.tex
\begin{tabular*}{\textwidth}{@{\extracolsep{\fill}}l rr rr rr@{}}
\toprule
 & \multicolumn{2}{c}{\textbf{required items seen}} & \multicolumn{2}{c}{\textbf{seen only by}} & \multicolumn{2}{c}{\textbf{WC003 score}} \\
\cmidrule(lr){2-3}\cmidrule(lr){4-5}\cmidrule(lr){6-7}
\textbf{world} & navigated & snapshots & navigated & snapshots & navigated & snapshots \\
\midrule
\tsixrows
\bottomrule
\end{tabular*}

%% file: figures/tab_mutation_body.tex
\begin{tabular*}{\textwidth}{@{\extracolsep{\fill}}l l r rrr@{}}
\toprule
 & & \textbf{before} & \multicolumn{3}{c}{\textbf{after removal} \textcolor{mute}{(correct = 0)}} \\
\cmidrule(lr){3-3}\cmidrule(l){4-6}
\textbf{removed item} & \textbf{how} & \textcolor{mute}{real} & \textbf{code-only} & \textbf{snapshot-only} & \textbf{ours} \\
\midrule
\tsevenrows
\bottomrule
\multicolumn{6}{@{}l}{\footnotesize\textcolor{mute}{untouched items in the same \tsevenbiomes{} biomes whose verdict changed (judge noise): \tsevenuntouched}}
\end{tabular*}

%% file: figures/tab_leaderboard_body.tex
\begin{tabular*}{\textwidth}{@{\extracolsep{\fill}}l rrrrr r@{}}
\toprule
\textbf{model} & \textbf{WC001} & \textbf{WC002} & \textbf{WC003} & \textbf{WC004} & \textbf{WC005} & \textbf{total} \\
 & \textcolor{mute}{/40} & \textcolor{mute}{/20} & \textcolor{mute}{/100} & \textcolor{mute}{/100} & \textcolor{mute}{/20} & \textcolor{mute}{/280} \\
\midrule
\tfourrows
\bottomrule
\end{tabular*}

%% file: sections/06_limitations.tex
\section{Limitations}
\label{sec:limitations}

\paragraph{Unreached code still earns half.}
The code half asks whether a quote is real text in the source, not whether that
code ever runs. In the mutation test, three features whose functions were kept but
never called still earned their code half, and a feature switched off with
\texttt{if(false)} was quoted by its disabled line. Checking that quoted code is
reachable from the world's build or frame loop would close most of this gap.

\paragraph{The visual judge can still be fooled.}
Verification narrows what the visual judge decides but does not remove it. Where no
pixel measurement applies, a look-alike can pass: drifting snowfall near the peaks
was accepted as a blizzard after the blizzard had been removed.

\paragraph{Placement is a model's reading.}
WC002's neighbour list comes from a model reading the layout code, and unlike the
item quotes it is not checked pair by pair. Against the actual biome maps of four
worlds it is right on 61 of 76 rule-relevant pairs (of 19 each:
\texttt{claude-fable-5-1} 17, \texttt{kimi-k-3} 14, \texttt{gemini-3-1-pro} 14, and 16
on \texttt{claude-opus-5-5}, a world outside the five reported here). The other two
reported worlds, including \texttt{grok-4-7}, which tops placement, have no such map,
so their placement scores are unchecked. The errors are mostly missed links, so a world can lose a placement point for a border it does have.
Reading each world's own biome map from the running page would make this exact, but
the code that stores that map differs in every world. In a trial on the same four
worlds, a model asked to locate it reproduced the exact map on its first attempt in
two; in the other two it missed a biome or returned values it could not map.

\paragraph{E2 (unobserved) mixes two causes.}
An item the frames do not show is either not rendered or not captured. The navigator
confirmed all ten biomes in three worlds, nine in one, and five in
\texttt{grok-4-7}'s, so part of that world's E2 points reflect capture, not
the world.

\paragraph{Noise and scale.}
We grade one world per model, from one prompt, once, and single item verdicts are
noisy (Section~\ref{sec:mutation}). Totals are steadier, but we treat gaps of a few points between neighbouring models
as ties and do not claim a ranking among them. The mutation test covers five
removals in one world. Repeated generations and repeated grading would give the
variance this paper does not report.

\paragraph{No human labels, by design.}
The worlds are open-ended and synthetic, with no reference answer to annotate, so
we validate the judge by construction rather than by agreement with people.
Mutation shows whether the judge notices what we changed, not whether it agrees
with a person on everything we did not.

\paragraph{Judges and medium.}
Every judge role is a model from the same developer as one model under
test (Appendix~\ref{app:judges}). That model ranked last, but we did not measure the
bias with a cross-family judge. The benchmark covers Three.js worlds only.

\section{Conclusion}
\label{sec:conclusion}

A world written as code has to be graded twice, once for what the program says and
once for what it shows, because the two disagree on a third of items, mostly code that no frame shows, and
neither can say alone which is right. \WB{} grades both: it explores the running
world close and wide, reads the code for what it is looking at, and lets an item
score only for evidence that survives a check against the source or the pixels.
Placement is the one reading we do not verify item by item; on four worlds it
matched the biome map on 61 of 76 pairs (Section~\ref{sec:limitations}).
Removing features on purpose shows the gain over reading alone, and the gap that
remains: code that exists but never runs.

%% file: figures/prompt_verbatim.tex
\begin{Verbatim}[breaklines=true,breakanywhere=false]
Build a single, self-contained HTML file that renders the world described
below as an explorable 3D scene using Three.js (CDN import is fine -- no
build step, no separate asset files). It must just open in a browser and
run, with a free camera (e.g. OrbitControls) so a visitor can orbit, zoom,
and pan around the whole world.

Include a simple on-screen UI so a visitor can navigate and read the
world easily (see UI section below).

Output ONLY the complete HTML file -- no prose, no commentary, no code fences.

Choose your own overall scale. What matters is relative placement,
ecological causality, and clear visual reading from orbit -- not any
particular absolute size or fixed elevation numbers.

--------------------------------------------------------
THE WORLD
--------------------------------------------------------

# Floating Biome Island with Ocean -- World Description

## Visual style (required)

Minecraft-like voxel / block aesthetic throughout:

-   Terrain, cliffs, water, lava, trees, flora, and structures are built
    from chunky axis-aligned cubes and blocky stacked forms -- not smooth
    organic meshes, not photoreal terrain, not low-poly "soft" stylization.
-   Flat, solid face colors (Lambert / unlit-friendly); no PBR gloss, no
    detailed textures required. Pixel-chunk readability from orbit.
-   Trees and plants are stacked blocks / simple voxel silhouettes
    (trunk + canopy cubes), not billboards or smooth cones.
-   Weather and particles can be simple quads or small cubes, but the
    landmass itself must read as a voxel island.

## Overview

A floating landmass hangs motionless in an infinite black void.
Viewed from a high orbit, nearly the entire surface should be readable at
once. Each biome needs room to read as its own region -- distinct
transitions, not a muddled cluster. Flora and fauna should feel alive
inside each biome without overcrowding the whole island.

Much of the outer perimeter is bordered by shallow coastal ocean and sandy
beaches before terrain rises inland. Elevations are exaggerated and
dramatic relative to each other: peaks dominate, shelves sit mid-height,
plains are low and open, cliffs drop sharply. The ocean is suspended with
the island and ends at cliff edges where waterfalls plunge into darkness.

## World layout (relationships)

Compose a coherent ecological sequence. Relative placement should feel
like this (adapt freely as long as the relationships hold):

1.  Snow mountains occupy the far / behind end of the island -- the final
    alpine massif at the cold edge.
2.  Snowy conifer forest sits as elevated foothills in front of those
    mountains (between the peaks and the rest of the island).
3.  Highlands occupy an elevated plain to the side of the snow forest --
    between the mountain flank and the back side of the flowering grove --
    then slope gently down into that pink grove.
4.  Dense jungle is a large wet canopy region, with a backwater / swamp
    marsh on the jungle's far side (away from the main plains outlet).
5.  Flat grassland plateau is the open river plain that receives wet-
    corridor drainage and leads toward the coast.
6.  Flowering grove is a moist pink basin off the plains / below the
    highland slope -- visually separate from green grassland.
7.  Desert is a vast arid basin (south / rain-shadow side) with its own
    sandstone desert mountains -- never the volcano.
8.  Volcano is a separate tall cone on volcanic ground only (far from
    the desert's river path).
9.  Coastal ocean and beaches wrap much of the southern and western edge.

## Scale and density (qualitative)

-   Prefer a clearly legible island: biomes spaced so each reads as its
    own region from high orbit.
-   Avoid collapsing adjacent biomes -- especially grassland vs flowering
    grove, desert vs volcano, snow forest vs alpine peaks.
-   Within each biome, flora and fauna should feel abundant for that
    climate (dense jungle canopy; cactus across desert flats; pines in
    the snow forest; sparse scrub on highlands).
-   Mountains and the volcano dominate the skyline; desert flats and the
    grassland plateau sit lower between them.

## Fauna

Voxel / blocky animals (stacked cubes, same aesthetic as the land) --
simple silhouettes are enough; place them where they belong:

-   Deer or similar grazers in jungle edges, grasslands, grove, conifers
-   Foxes or small arid / marsh creatures in desert and swamp
-   Occasional coastal or river life near beaches and deltas if it fits
-   Enough motion or presence to feel alive, not empty dioramas --
    without overcrowding

## Ecological reasoning

Layout and systems must follow ecological logic, not decoration:

-   Water stays in the wet corridor: mountain melt -> through / past the
    snow-forest foothills and wet highlands -> weave through jungle ->
    flat grasslands -> coastal delta / ocean.
-   Flowering grove is a moist offshoot below the highland slope / plains
    moisture -- not a stop on the way into the desert.
-   Desert is arid: vast flat basin with sandstone desert mountains;
    cacti across the flats; only fading dry washes; no through-rivers
    into the volcano; no perennial river cutting the desert core.
-   Volcano is a separate tall cone with lava tributaries and lava pools
    on volcanic ground. Where water meets lava, rock quenches to
    obsidian.
-   Rain falls downward from moving cloud decks over wet biomes.
-   Snowfall comes from cloud decks over the snowy conifer forest.
-   Blizzard / white windstorm lives on the behind face of the snow
    mountains (the far alpine side), not as a front-slope decoration
    facing the forest alone.
-   Snowy trees carry snow on their canopies, not only on the ground.
-   Highlands are genuinely elevated relative to plains and grove, with
    a gentle downslope into the pink forest.

## Biomes

### 1. Snow Mountain Range (far / behind)

-   Tall alpine peaks -- high relief, not gentle hills
-   Snow on summits and upper ridges; rock on lower slopes
-   Meltwater feeding the wet corridor
-   Active blizzard on the behind mountain face: fast whitish wind
    sheets (sandstorm motion, white instead of brown), spanning a wide
    strip of that rear alpine tape

### 2. Snowy Conifer Forest (foothills in front of the peaks)

-   Elevated shelf / foothill band in front of the snow mountains
-   Dense pines with snow on canopies as well as ground
-   Cloud deck directly above the forest that drops snowfall onto it
-   Distinct from both bare alpine rock and lowland plains

### 3. Temperate Highlands (elevated plain beside the snow forest)

-   Sit to the side of the snow forest, between mountain influence and
    the back of the flowering grove -- an elevated open highland plain
-   Higher than grasslands and grove; rocky hills, scrub, hardy plants
-   Gentle slope down into the pink flowering grove
-   Occasional mist and light rain

### 4. Dense Jungle / Rainforest

-   Thick emerald canopy, layered undergrowth, densely packed trees
-   Rocky clearings, inland ponds, major waterfall
-   On the jungle's backside (away from the plains outlet): a backwater
    / swamp ecosystem -- murky standing pools, mangroves or similar,
    low mist, stagnant channels (not a desert river)
-   Persistent jungle rain: rainfall shafts, wet canopy, rising mist
    after showers

### 5. Central Grasslands (flat river plateau)

-   Open flat green plain -- clearly not the pink grove
-   Dense winding river network with many tributaries
-   Scattered woodland copses and yellow/white wildflowers
-   Meadow color green/olive, not pink

### 6. Flowering Grove

-   Moist basin receiving the highland downslope and/or plains moisture
-   Dense pink flowering canopy and pink-tinted meadow floor
-   Small streams; visually different from grassland at a glance

### 7. Desert

-   Mostly flat arid land
-   Distinct desert mountains / sandstone peaks on that flat floor --
    these are NOT the volcano
-   Dry washes that fade out; no perennial river through desert core
    or into the volcano
-   Cacti and life spread across the flats
-   Desert winds and sandstorm / dust storm

### 8. Volcanic Wasteland

-   Tall stratovolcano cone with glowing crater -- not a flat puddle
-   Lava tributaries into lava pools on volcanic ground only
-   Black basalt, ash plumes, periodic eruptions
-   Water + lava contact becomes obsidian

## Coastal Ocean

Southern and western edges surrounded by turquoise coastal waters:

-   Shallow lagoons and deeper offshore water
-   Rocky shoreline, small islands and reefs
-   Multiple river mouths and a visible braided delta into the ocean
-   Freshwater mixing into saltwater at the mouths

## Beaches

Natural transition from grasslands and forests into beaches:

-   Bright sand, curved bays, warm-region palms where fitting
-   Rocky headlands, clear shallow water, coastal vegetation

## Rivers and Hydrology

A rich branching system that obeys ecological law:

-   Mountain melt and highland moisture feed headwaters.
-   Flow weaves the wet corridor through jungle into the flat grassland
    plateau, then widens into a coastal delta.
-   Grove receives a moist feeder from highland slope and/or plains --
    not a desert route.
-   Desert stays dry aside from fading washes; no river into the volcano.
-   Swamp holds stagnant backwater connected to the jungle wet side.
-   Show the water cycle: rain and melt feed rivers; heat drives
    evaporation mist; moisture returns as rain or snow.

## Climate, Clouds, Seasons, and Sun

-   Rain, snow, mist, sandstorm, ash, and cloud decks arrive in cyclic
    bursts -- fade in, hold, fade out, quiet gap -- staggered by biome.
-   Snowfall: cloud deck over the conifer foothill forest.
-   Blizzard: whitish high-altitude wind sheets on the behind alpine
    face of the snow mountains.
-   Desert: winds and sandstorm. Volcano: eruptions, ash, lava pools;
    water+lava -> obsidian.
-   Reliable seasonal cycle (spring -> summer -> autumn -> winter)
    modulating rain/snow intensity and light tint.
-   Visible square sun on a cyclic day / night pattern: bright noon,
    dimmer dusk and night (the void stays black; the land darkens).

## UI (required for easy navigation)

Keep the HUD minimal and readable over the black void:

-   Biome legend: color swatch + biome name for every major region
    (mountains, snow forest, highlands, jungle, swamp, grasslands,
    flowering grove, desert, volcano, coast / ocean). Optionally note
    each biome's weather flavor (rain, blizzard, sandstorm, etc.).
-   Click (or tap) a legend entry to fly / animate the camera to that
    biome; include a "whole island" overview entry that returns to a
    comfortable high orbit.
-   Season readout: show the current season name on screen and update
    it as the seasonal cycle advances.
-   Brief control hint in the legend footer (e.g. orbit / zoom / pan).

## Cliffs and Underside

-   Vertical rock walls around the perimeter
-   Tall stone foundation matching heightened elevations above
-   Numerous waterfalls; floating rock fragments beneath
-   Blue glowing mineral seams in places

## Color Palette

  Region       Colors
  ------------ -----------------------------
  Ocean        Turquoise, deep blue, cyan
  Beaches      White, cream, tan
  Jungle       Emerald, dark green, mist grey
  Swamp        Murky olive, dark water
  Plains       Green, olive
  Grove        Pink blossoms, soft meadow
  Highlands    Grey-green scrub
  Mountains    White summits, blue-grey rock
  Conifers     Dark pine, snow-capped canopy
  Desert       Gold, ochre, dust haze
  Volcano      Black, orange, crimson, ash
  Rivers       Cyan (murkier in swamp)
  Weather      Rain blue, snow white, sand tan, ash grey

## Lighting

-   Square sun drives a full day -> night cycle; seasonal tint modulates
    warmth
-   Nights are clearly darker than noon (ambient + sun intensity drop)
-   Local weather tinting (cooler in storms, ember glow near volcano)
-   Crisp terrain shadows when weather is clear
-   High visibility of the whole island from a comfortable orbit by day
-   Black void background with no stars or atmosphere dome

## Visual Composition

A Minecraft-style voxel floating continent: far snow mountains with a
behind-face blizzard; elevated snow-forest foothills under snowfall
clouds; highlands on an elevated plain sloping into a pink grove; a large
dense jungle with a backwater swamp on its far side; a flat grassland
river plateau; arid desert with its own sandstone mountains separated
from a tall erupting volcano; wet-corridor rivers to a coastal delta;
cyclic weather; square sun with day/night and seasons; clickable biome
legend and season HUD. The ocean borders the land before plunging over
cliff edges into the void.

Success is ecological clarity, vivid living systems, easy navigation,
and a coherent blocky voxel look -- invent your own scale; make it as
strong or stronger as this description allows.
\end{Verbatim}

%% file: sections/E_scoring.tex
\section{Scoring details}
\label{app:scoring}

\paragraph{Per test.}
WC003 and WC004 follow Eqs.~\ref{eq:credit}--\ref{eq:biome}. WC005 applies
Eq.~\ref{eq:credit} to each of its items, except weather items, which are scored from
code alone. The other two tests are
\begin{equation*}
\mathrm{WC001} = \sum_i p_i\,\pi_i\,[\text{quote verified}_i] \;+\; 20\cdot\frac{n_{\text{clean}}}{n_{\text{judged}}},
\end{equation*}
where $\pi_i$ is the source judge's probability for item $i$ (its points total 20) and
the second term is the share of captured views in which the visual bug hunt found no
defect, and
\begin{equation*}
\mathrm{WC002} = \sum_{b=1}^{10}\Bigl([\text{covered}_b] + [\text{covered}_b \land \text{rules}_b]\Bigr),
\end{equation*}
one point for a biome whose layout code is found and verified, and one more if its
placement rules pass. The total is $\sum_{k=1}^{5}\mathrm{WC}_k \le 280$. Two rules act on the
delta alone. In WC003 its score is zero when the frames show the ocean running out over
the void with no blocks beneath. WC004 drops its delta points when WC001's bug hunt sees
water over the void in at least two wide views; the gate fires on one reported world,
\texttt{kimi-k-3}.

\paragraph{Pixel measures.}
Table~\ref{tab:pixels} defines the measurements that decide instead of the visual
judge wherever a property is measurable.

\begin{table}[!htb]
  \centering\small
  \begin{tabularx}{\textwidth}{@{}l Y Y@{}}
  \toprule
  \textbf{measure} & \textbf{definition} & \textbf{pass condition} \\
  \midrule
  brightness $L$ & mean grey level of non-void pixels (grey $>12$), left 20\% of
    the frame (the legend) cropped
    & day $=\arg\max_t L(t)$; a day/night cycle exists if $L_{\text{day}}-L_{\text{night}}\ge 4$ \\
  chromaticity & $(r,b)$: red and blue shares of land pixels
    & shift $=\max(\lvert\Delta r\rvert,\lvert\Delta b\rvert)\ge 0.015$ (seasons), $\ge 0.03$ (twilight) \\
  light change & mean per-channel difference from the day frame & $\ge 8$ grey levels \\
  motion & share $f$ of pixels whose grey level changes by more than 24 within a burst & $f \ge 0.001$ \\
  \bottomrule
  \end{tabularx}
  \caption{Pixel measures and their thresholds, as implemented in the capture and
  grading code.}
  \label{tab:pixels}
\end{table}

%% file: sections/B_breakdown.tex
\section{Score breakdown}
\label{app:breakdown}

The leaderboard (Table~\ref{tab:leaderboard}) gives one number per test. This
appendix splits each test into the parts it is made of, for every model. Colours
follow Figure~\ref{fig:scores}.

\begin{figure}[!htb]
  \centering
  \includegraphics[width=\textwidth]{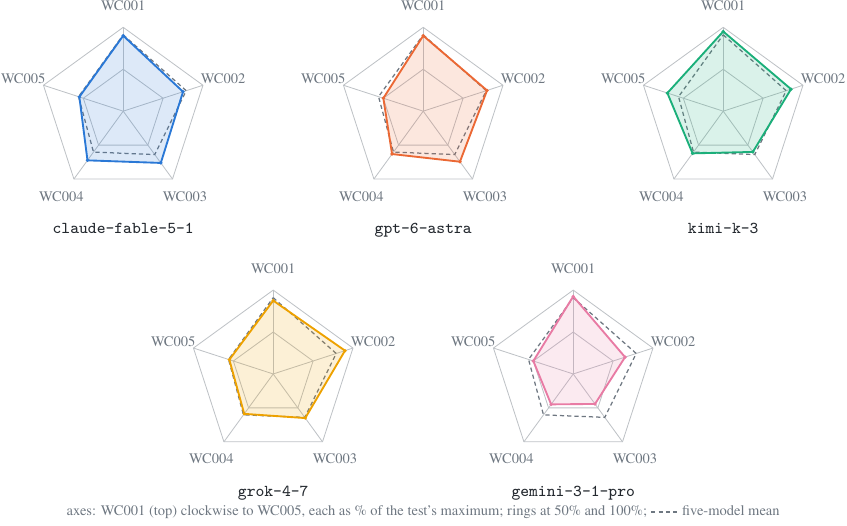}
  \caption{Score profile per model: each axis is one test (WC001 at the top,
  clockwise to WC005) as a share of its maximum; the dashed polygon is the
  five-model mean.}
  \label{fig:radar}
\end{figure}

\paragraph{Profile across tests.}
Figure~\ref{fig:radar} draws each model's five test scores, each as a share of the
test's maximum, against the five-model mean. Models differ most on
\bdSpreadWideTest{} (a \bdSpreadWide-point spread in percentage terms between the
best and worst model) and least on \bdSpreadNarrowTest{} (\bdSpreadNarrow{} points):
every model builds a recognisable voxel island, and they separate on what the
island contains and how it is laid out. We show one radar per model rather than all
five overlaid because, overlaid, two models that tie on two tests hide each other.

\begin{figure}[!htb]
  \centering
  \includegraphics[width=\textwidth]{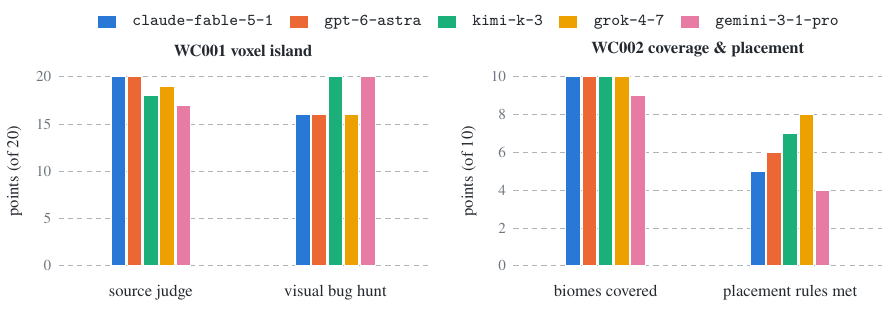}
  \caption{WC001 and WC002 split into their parts. WC001: source judge and visual bug
  hunt, 20 points each. WC002: one point per covered biome and one per biome whose
  placement rules pass, 10 each.}
  \label{fig:bdchecks}
\end{figure}

\paragraph{WC001 and WC002 by part.}
WC001 is two checks of 20 points each (Figure~\ref{fig:bdchecks}, left): the source
judge scores \bdSrcMin--\bdSrcMax{} and the visual bug hunt \bdBugMin--\bdBugMax{}
across models, so the voxel test is close to saturated. WC002 earns one point per
biome covered and one per biome whose placement rules pass (right). Coverage is
near complete (\bdCovMin--\bdCovMax{} of 10 biomes), while placement rules range
from \bdRuleMin{} to \bdRuleMax{} of 10: nearly every biome is built, but not where
the prompt says. The one uncovered biome is \texttt{gemini-3-1-pro}'s grove: its
layout quote did not verify against the source, and no blind-confirmed frame shows it.

\begin{figure}[!htb]
  \centering
  \includegraphics[width=\textwidth]{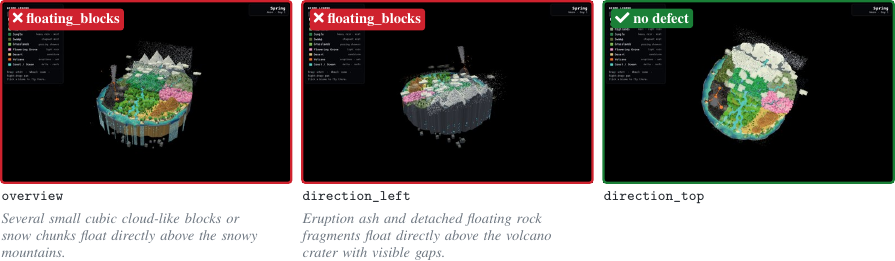}
  \caption{WC001 visual bug hunt, reference world: two of the three flagged
  views (the third, a delta close view, is not shown) and one clean view, with the
  judge's own words for each flag. Both flagged views show things the prompt requires:
  cloud decks, ash, and floating rock fragments beneath the island.}
  \label{fig:wc001}
\end{figure}

\paragraph{WC001 in one world.}
Figure~\ref{fig:wc001} shows WC001's two halves on the reference world. The visual
bug hunt looks at each captured view on its own and flags floating blocks, hollow
masses or non-voxel shapes; it flagged \wcaBuggy{} of \wcaJudged{} views (the overview, the left view and a
close view of the delta), for \wcaBugScore{} of 20; the figure shows the first two.
The flags on the two wide views charge the world for things the prompt asks for.
The prompt requires cloud decks, ash plumes and ``floating rock fragments beneath''
the island, and this world builds all three, the fragments as 70 deliberately
detached cubes. The overview is flagged for ``cubic cloud-like blocks'' above the
peaks and, in most of its votes, for detached blocks under the island. The left view
is flagged for eruption ash and, again, for blocks under the island. The top view is marked clean, so the same world passes or fails
by camera angle. The third flag, flat slabs in the air near the volcano in the delta close view, we did not
trace. The bug hunt cannot tell a required floating feature from a defect, so it
can cost a faithful world points. Water over the void is counted once per world,
and only when at least two views show it. The source judge (Table~\ref{tab:wc001})
gives each of its items a probability and a quote; an item earns
$p_i\,\pi_i$ only if the quote verifies against the source, so it cannot talk a
world into points. Here every item verifies, for \wcaSrcScore{} of 20.

\begin{table}[!htb]
  \centering\small
  \begin{tabular*}{\textwidth}{@{\extracolsep{\fill}}l r r c r l@{}}
  \toprule
  \textbf{item} & \textbf{points} & \textbf{prob.} & \textbf{verified} & \textbf{earned} & \textbf{quote (start)} \\
  \midrule
  \wcaRows
  \bottomrule
  \end{tabular*}
  \caption{WC001 source judge, reference world: each item's points, the judge's
  probability, whether its quote is text the source contains, and the points earned.}
  \label{tab:wc001}
\end{table}

\begin{figure}[!htb]
  \centering
  \includegraphics[width=\textwidth]{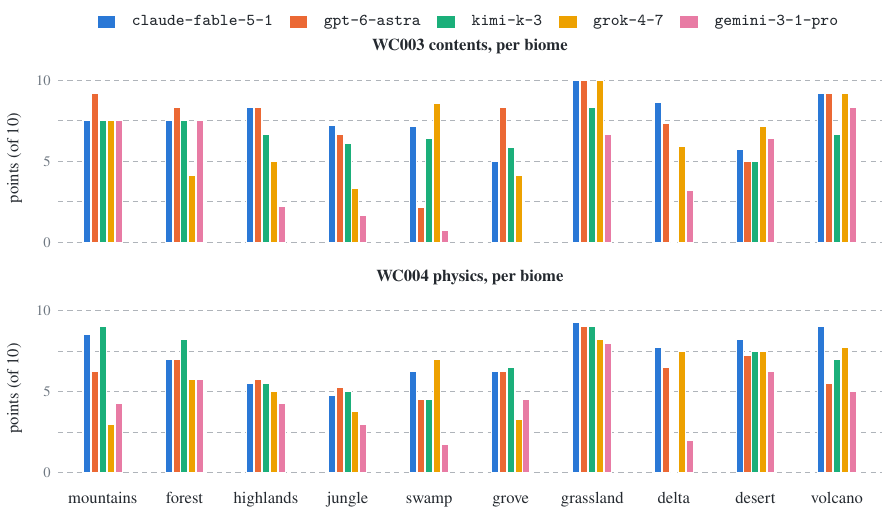}
  \caption{WC003 (contents) and WC004 (physics) per biome, 10 points per biome in
  each test, after the island gate (which zeroes \texttt{kimi-k-3}'s WC004 delta). A missing
  bar is a score of zero.}
  \label{fig:bdbiomes}
\end{figure}

\paragraph{WC003 and WC004 by biome.}
Each biome is worth 10 points in each of the two large tests
(Figure~\ref{fig:bdbiomes}). In WC003 the lowest five-model average is
\bdContentsLo{} (\bdContentsLoAvg{} of 10) and the highest \bdContentsHi{}
(\bdContentsHiAvg); in WC004 the lowest is \bdPhysicsLo{} (\bdPhysicsLoAvg) and the
highest \bdPhysicsHi{} (\bdPhysicsHiAvg). The per-biome bars show what the averages
hide: in several biomes one model scores near the maximum while another scores near
zero.

\begin{figure}[!htb]
  \centering
  \includegraphics[width=\textwidth]{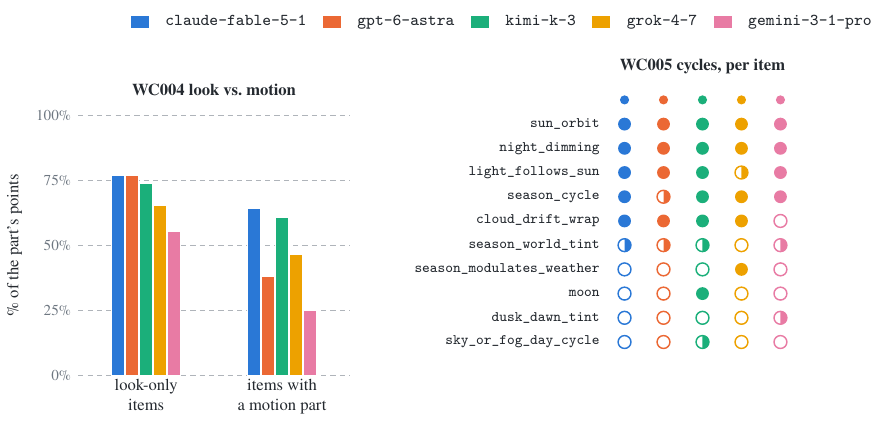}
  \caption{\emph{Left}: WC004 split into look-only items and items with a motion part,
  as a share of each part's points. \emph{Right}: WC005 per item and model: filled =
  full credit, half-filled = half, ring = none. Rows are ordered by total credit
  across models.}
  \label{fig:bdcycles}
\end{figure}

\paragraph{WC004 look and motion; WC005 by item.}
WC004 mixes items that are judged on appearance alone (\bdLookPts{} points per world)
with items that also carry a motion part (\bdMotionPts{} points).
Over all five worlds, models earn \bdLookPct\% of the look-only points but
\bdMotionPct\% of the points on items with a motion part (Figure~\ref{fig:bdcycles},
left): things are more often built than made to move. WC005 has \bdWcItems{} items
(right). In every world, \bdWcFullN{} of them earn full credit and \bdWcZeroN{}
earn nothing; the rest split the models. The three items with the least
credit across models are \bdWcLowList.

%% file: sections/C_failures.tex
\section{Design history: judges that failed}
\label{app:failures}

The judge in Section~\ref{sec:method} was not designed up front. Each of its checks
replaced a simpler judge that was built first, run on real worlds, and found to
fail in a specific, measurable way. Table~\ref{tab:failures} lists them. Two
patterns repeat. Judges that match the source text (keywords, patterns) reward what
a program \emph{names}, so they are easy to game and brittle under harmless
rewrites. Judges that trust a model's answer (a quote, a sighting, an ``it changed'')
accept fluent claims that are not true. Every replacement therefore anchors the
claim to something the model under test cannot fake: the source text itself, or
measured pixels.

\begin{table}[!htb]
  \centering\small
  \input{figures/tab_failures_body.tex}
  \caption{Simpler judges built first, how each failed on real worlds, and what
  replaced it.}
  \label{tab:failures}
\end{table}

%% file: figures/tab_failures_body.tex
\begin{tabularx}{\textwidth}{@{}>{\hsize=.75\hsize}Y >{\hsize=1.25\hsize}Y Y@{}}
\toprule
\textbf{simpler judge} & \textbf{how it failed} & \textbf{replaced by} \\
\midrule
keyword biome coverage
  & every scored model got 10/10; a one-line fake, \texttt{const pine=1,\dots,lava=9}, also got 10/10
  & a biome counts only if its placement code is quoted and found in the source \\
regex voxel check (83 patterns)
  & 43 patterns matched exactly one model's file; renaming variables without changing behaviour moved scores by up to 14/38
  & code judge giving a probability and a verbatim quote per item \\
unchecked code quotes
  & a plausible but invented \texttt{scene.add(new THREE.Mesh(palmGeo, \dots))} scored
  & a quote counts only if it is text the source contains \\
strict quote check alone
  & real code joined onto one line or shortened with ``\dots'' was rejected: sun orbit, season cycle and moon lost (14 $\to$ 8)
  & repair that picks source line numbers and rebuilds the quote from the file (8 $\to$ 13) \\
asking the VLM ``does this show the swamp?''
  & said yes on all 30 frames across three worlds, including a volcano saved as the swamp
  & blind identification of each frame before any item is judged \\
VLM judging lighting change
  & called frames ``identical'' whose mean brightness was 11.9 / 35.0 / 11.6 / 11.5
  & change measured from pixels: brightness, frame difference, chromaticity \\
trusting the clock patch
  & one world's noon landed on the wrong requested time
  & the brightest of four renders is taken as day \\
\bottomrule
\end{tabularx}

%% file: sections/A_judges.tex
\section{Judge models and settings}
\label{app:judges}

The main text says ``the judge'' for readability. Several models fill distinct
roles, and every role is set in one configuration file, so a run can be repeated
with the same judges or a whole judge set swapped (Table~\ref{tab:judges}).

\begin{table}[!htb]
\centering
\small
\begin{tabularx}{\textwidth}{@{}l Y l@{}}
\toprule
\textbf{role} & \textbf{what it does} & \textbf{model} \\
\midrule
vision & item visibility in frames; blind biome identification & \texttt{gemini-3.5-flash-lite} \\
code probe & quotes the code that builds each item (WC003--WC005) & \texttt{gemini-3.1-flash-lite} \\
text judge & WC001 source judge, clock patch, evidence agent & \texttt{gemini-3.1-flash-lite} \\
navigator & drives the camera to each biome & \texttt{gemma-4-26b-a4b-it} \\
extract & WC002 biome slicing; layout reading from the whole source & \texttt{gemma-4-31b-it} \\
quote repair & picks line numbers for a badly copied quote & \texttt{gemma-4-26b-a4b-it} \\
\bottomrule
\end{tabularx}
\caption{Judge roles. A request still failing after three retries on a server
error falls back along \texttt{gemma-4-31b-it} $\rightarrow$
\texttt{gemma-4-26b-a4b-it} $\rightarrow$ \texttt{gemini-3.1-flash-lite}; a call
larger than a model's per-minute token cap is sent to the first fallback whose cap
fits. WC002's classification and layout reading both send the whole source, which for
every reported world but \texttt{gemini-3-1-pro}'s is larger than the Gemma caps, so
those calls land on the last.}
\label{tab:judges}
\end{table}

\paragraph{Settings.} Frames are sent to the vision judge at 896\,px wide and to the
navigator at 640\,px. The navigator has 8 actions per biome; the evidence agent has
10 tool calls per search. Output is capped at 16{,}384 tokens per call, and each
request times out after 300\,s.

\paragraph{Shared family with a model under test.} All six judge roles are Gemini or Gemma
models, the same developer as one of the five models under test,
\texttt{gemini-3-1-pro}. Judges are known to favour outputs from their own family \citep{panickssery2024self}.
The one comparison we have does not show favouritism: \texttt{gemini-3-1-pro} ranks
last overall (149.00 of 280) and last on WC003 and WC004, the tests that rely most on
the vision and code-probe judges. That is a single model, and a low score does not
rule out a smaller bias. We did not run a cross-family judge to measure the effect
directly.